\documentclass{article}
\usepackage[a4paper, total={6in, 9in}]{geometry}
\usepackage[english]{babel}
\usepackage[utf8]{inputenc}
\usepackage[T1]{fontenc}
\usepackage{setspace}
\usepackage{float}
\usepackage{amsmath}
\usepackage{amssymb}
\usepackage{amsthm}
\usepackage[title]{appendix}
\usepackage{comment}
\usepackage{csquotes}
\usepackage{pdfpages}
\usepackage{lmodern}
\usepackage[most]{tcolorbox}
\usepackage{blindtext}
\usepackage{hyperref}
\usepackage{mathtools}
\usepackage{array}
\newcolumntype{C}{>{\centering\arraybackslash}p{1.5cm}}
\usepackage[format=hang, font=it]{caption}
\newtheorem{Theorem}{Theorem}[section]
\newtheorem{Definition}[Theorem]{Definition}
\newtheorem{Remark}[Theorem]{Remark}

\renewcommand{\epsilon}{\varepsilon}

\usepackage[backend=biber, maxbibnames=10, maxcitenames=2, style=authoryear, natbib]{biblatex}
\title{Symmetric positive semi-definite Fourier estimator of instantaneous variance-covariance matrix}
\date{\today}
\author{Jir\^o Akahori\thanks{Ritsumeikan University}, Nien-Lin Liu\thanks{School of Management, Tokyo University of Science, e-mail:nl-liu@rs.tus.ac.jp}, Maria Elvira Mancino\thanks{Dept. of Economics and Management, University of Florence},  Tommaso Mariotti\thanks{Scuola Normale Superiore, Pisa, e-mail: tommaso.mariotti@sns.it}, Yukie Yasuda\footnotemark[1]}

\begin{document}
\maketitle
\begin{abstract}
In this paper we propose an estimator of spot covariance matrix which ensure symmetric positive semi-definite estimations. The proposed estimator relies on a suitable modification of the Fourier covariance estimator in \citet{malliavin2009fourier} and it is consistent for suitable choices of the weighting kernel. The accuracy and the ability of the estimator to produce positive semi-definite covariance matrices is evaluated with an extensive numerical study, in comparison with the competitors present in the literature. The results of the simulation study are confirmed under many scenarios, that consider the dimensionality of the problem, the asynchronicity of data and the presence of several specification of market microstructure noise.
\end{abstract}

\section{Introduction}\label{sec:intro}

Empirical studies have pointed out the importance of considering distinct time variations in correlations between asset prices. Then, in the last years,
several studies have addressed the issue of efficiently estimating covariances
using high frequency data asynchronously sampled across different assets.
While the literature is becoming rich as it concerns the estimation of integrated covariances, it is still sparse for the spot covariances estimation.
An early proposal to cope with spot covariances estimation with asynchronous high frequency data, has been given in \citet{malliavin2002fourier}. In contrast with the other estimators which rely on a pre-processing of data in order to make them
synchronous, such as linear interpolation, piecewise constant (previous-tick)
interpolation or the refresh-time procedure proposed by \citet{barndorff2011multivariate}, the Fourier estimator uses all the available data, being based on an integration procedure. The possibility of using all data, avoiding any preliminary manipulation of them (such as pre-averaging, see, e.g., \citet{ait2014high}), translates into the direct use of unevenly sampled
returns and even asynchronous data in the multivariate case. The estimator proposed by \citet{hayashi2005covariance} proposes to circumvent the drawbacks caused by asynchronicity by considering the contribution of the product of the price increments only if the corresponding observation intervals are overlapping. The Pre-averaged All-Overlapping estimator by \citet{christensen2010pre} builds on the same idea, with the aim to render the estimator by \citet{hayashi2005covariance} robust to microstructure noise.

From a practical point of view, the choice of which estimators to use should not be only based on the rate of convergence to their asymptotic distributions. For instance, the fact that the estimated covariance matrix preserves its symmetry and positive semi-definiteness is a primary issue.
This property has important consequences in several contexts, such as
the recently developed field of principal component analysis with high-frequency data (\citet{liu2017approximation}, \citet{ait2019principal}, \citet{chen2020five}) or the asset
allocation framework (see, e.g., \citet{engle2006testing}).
While this point has been addressed by some authors for the integrated covariances estimators (see, e.g, \citet{barndorff2011multivariate}, \citet{mancino2011estimating} \citet{park2016estimating}, \citet{cui2019large}), at the best of our knowledge the estimator of spot covariances
proposed in the present paper is the first to guarantee positive semi-definiteness of the estimation itself, a problem that so far has not been addressed in the literature. For example, when dealing with spot volatility, \citet{chen2020five} integrate the estimations before computing the eigenvalues of the covariance matrix, while in \citet{bu2022nonparametric} positive semi-definiteness is imposed applying suitable shrinkage techniques to the estimation, thus introducing a manipulation of the estimated matrix.

The aim of this work is to propose a novel spot covariance estimator which is symmetric and positive-definite and analyze its finite-sample properties in a simulated environment. Our starting point is the spot Fourier estimator by \citet{malliavin2009fourier}. Also this estimator, however, due to lack of symmetry in the Fej\`{e}r kernel, may fail to provide positive semi-definite estimations when the asset prices are observed on asynchronous grids. To guarantee that the estimations are symmetric and positive semi-definite, in this paper we introduce a modified version of the Fourier estimator, which we call Positive-Definite Fourier (hereafter PDF) estimator, and we prove that indeed fulfill the desired property.

The proposed estimator relies on two parameters: the cutting frequency $N$, and the localization $M$. The question of how to optimally choose them in order to minimize the error is assessed via a simulation study, where a grid of possible values is tested against several different model specification for both the efficient price process and the additive microstructure component. We find that the optimal cutting frequency $N$ is stable between different models for the volatility process, and in absence of noise $N=\frac{1}{2}n^{3/4}$ should be chosen, while in presence of noise a value of $N=\frac{1}{2}n^{2/3}$ is more suitable, in line with the other Fourier-type estimators. In both cases, choosing $N < n/2$, i.e. smaller than the Nyquist frequency, leads to a reduced error, while reducing also the computational effort required by the estimator. The parameter $M$ seems not to be significantly affected by microstructure noise, and the optimal value appears to be $M=N^{4/9}$ or $M=N^{1/2}$ depending on the model for volatility, with small differences between the two choices for the cases analyzed in our study.

In addition to that, the sensitivity of $N$ is further studied in a dedicated simulation exercise, where we specifically analyze the impact of introducing asynchronicity, in comparison to a scenario with synchronous sampling. The above-mentioned choice of $N$ is confirmed as preferable to the Nyquist frequency, since when the correlation between the two assets is smaller than 1 both the absolute bias and the mean square error increase for large values of $N$ in presence of asynchronicity, showing however that exists an interval of values of $N$ that emerges substantially unaffected.

Thereafter, to evaluate the finite-sample performance of the proposed PDF estimator, we compare its accuracy and the percentage of positive semi-definite estimations that it is able to produce with the ones obtained employing the smoothed two-scale estimator by \citet{mykland2019algebra} and the local method of moments estimator by \citet{bibinger2019estimating}, which are both able to manage asynchronous observations. Developing this comparison we focus on the main problems that may affect the estimation of variance-covariance matrices using high frequency data. First of all we consider a starting case in absence of market microstructure noise; secondly we focus on the level of asynchronicity, considering different intensities of the Poisson processes that drives the observation frequency; last we analyze the presence of market microstructure noise, considering noise coming from rounding, i.i.d noise, autocorrelated noise, noise correlated with the efficient price process and heteroskedastic noise. It is shown that, in this exercise, the PDF estimator is the only one to consistently produce positive semi-definite estimations in 100\% of the cases, while maintaining an hedge with respect to the competitors in terms of mean square error. Moreover, while the issue of dimensionality in the literature has been mainly considered in relation to PCA, through the comparison we consider whether increasing dimension may change the empirical distribution of the eigenvalues to the point of invalidate the ability of estimators to produce positive semi-definite matrices. As a result, the competing estimators are shown to produce a lower proportion of positive semi-definite matrices when the number of assets considered increases.

The robustness of all the simulation results are confirmed changing the simulation model behind the analysis; in particular we consider: an Heston Stochastic Volatility model \citet{heston1993closed}, a One Factor Volatiltiy model, a Two Factor Volatility model (\citet{chernov2003alternative}) and a Rough Heston model (\citet{el2019characteristic}), getting in each case comparable results.

The  remainder of this work is organized as follows. In section \ref{sec:est} the positive semi-definite (PDF) Fourier estimator of spot covariance is introduced, and its positivity is proved. Section \ref{sec:simul} contains the simulation study including a sensitivity analysis on the parameter of the proposed estimator in terms of mean square error of the estimation, and a comparison between the proposed estimator and alternative estimators present in the literature, in which accuracy and ability to produce positive semi-definite matrices are considered. Section \ref{sec:conc} concludes.

\section{The positive semi-definite estimator}\label{sec:est}
Let $X=(X^1, ..., X^d)$ be a $d$-dimensional It\^o semimartingale

\[
X^j_t = x^j_0 + \int_0^t b_j(s)ds + \sum_{k=1}^d \int_0^t \sigma_{jk}(s) dW_s^k, \quad j,k=1,...,d
\]
with $W=(W^1, ..., W^d)$ a $d$-dimensional Brownian motion on the filtered probability space $(\Omega, ({\cal F}_t)_{t\in [0,T]}, P)$ and $b_j$ and $\sigma_{j k}$ are adapted processes.

In this work, we are interested in the non-parametric estimation of the $d\times d$ instantaneous (spot) covariance matrix $V(t)$, with entries
\[
V^{j,j'}(t) \coloneqq \sum_{k=1}^d \sigma_{j k}(t) \sigma_{j' k}(t), \quad \text{for} \quad j,j'=1,...,d \quad \text{and} \quad t\in [0,T].
\]

We assume that the prices are observed on discrete, irregular and asynchronous time grids

\[
0=t^j_0 < t^j_1 < , ..., < t^j_{n_j}=2\pi \quad \text{ for } j=1, ..., d,
\]
where the assumption $T=2\pi$ is done for simplicity of notation and is not restrictive. In the following, $\Delta X^j_l$ denotes the discrete return $X^j_l-X^j_{l-1}$ for $j=1,...,d$ and $l=1,...,n_j$.

In this setting, we propose the following estimator of spot covariance.

\begin{Definition} Let $\mathcal{K}$ be a finite subset of $\mathbb{Z}$, $\mathcal{S} \coloneqq \{\mathcal{S}(k) \subset_{\text{finite}} \mathbb{Z}^2 \colon k \in \mathcal{K}, (s,s') \in \mathcal{S}(k) \implies s+ s' =k \}$, and $c$ be a complex function on $\mathcal{K}$; we define the estimator for {\bf $V^{j,j'}(t)$} as:
	
\begin{equation}\label{eq:def}
\hat{V}^{j,j'}_{\mathcal{K},\mathcal{S}}(t)= \sum_{l=1}^{n_j}\sum_{l'=1}^{n_{j'}}\sum_{k \in \mathcal{K}} c(k) e^{ikt}\sum_{(s,s') \in \mathcal{S}}e^{-ist^j_l}e^{is't^{j'}_{l'}} \Delta X^j_l \Delta X^{j'}_{l'}.
\end{equation}
\end{Definition}

\begin{Remark}
If we take $\mathcal{K}=\{0, \pm 1, \pm2, ..., \pm M\}$ for some positive integer $M$ and $\mathcal{S}(k)= \{(s,s')|s+s'=k, s=0, \pm 1, \pm2, ..., \pm N\}$ for some positive integer $N$, and
	\[
	c(k) = \frac{1}{2\pi}\left(1-\frac{|k|}{M+1}\right) \frac{1}{2N+1},
	\]
	we obtain:
	\[
	\hat{V}^{j,j'}_{\mathcal{K},\mathcal{S}}(t)=\sum_{k=-M}^M \left(1-\frac{|k|}{M+1}\right) e^{ikt}\sum_{s=-N}^N \sum_{l=1}^{n_j}\sum_{l'=1}^{n_{j'}} e^{-ist^j_l}e^{i(k-s)t^{j'}_{l'}} \Delta X^j_l \Delta X^{j'}_{l'}
	\]
	Using the Dirichlet and the Fej\`{e}r kernels defined as $D_N(x) =  \sum_{k=-N}^N e^{ikx}$ and $F_M(x) = \sum_{k=-M}^M \left(1-\frac{|k|}{M+1}\right)e^{ikx}$, the estimator can be written as follows:
	\[
	\hat{V}^{j,j'}_{\mathcal{K},\mathcal{S}}(t)=\frac{1}{2\pi}\frac{1}{2N+1}\sum_{l=1}^{n_j}\sum_{l'=1}^{n_{j'}}F_M(t-t^j_l)D_N(t^j_l-t^{j'}_{l'})\Delta X^j_l \Delta X^{j'}_{l'},
	\]
and the estimator (\ref{eq:def}) coincides with the Fourier spot covariance estimator introduced by \citet{malliavin2009fourier}, whose asymptotic properties have been studied in \citet{mancino2015fourier} (in the absence of noise) and in \citet{mancino2022asymptotic} (in the presence of noise). However, while the positivity of the corresponding estimator of the integrated covariance matrix is proved in \citet{mancino2011estimating}, the spot covariance estimator may fail in producing symmetric positive semi-definite estimations, being $F_M(t-t^j_l)D_N(t^j_l-t^{j'}_{l'})$ not symmetric in $j,j'$, leading to complex eigenvalues in $\hat{V}_{\mathcal{K},\mathcal{S}}(t)$.
\end{Remark}

The main theoretical result of this work concerns the positive semi-definiteness of the proposed estimator, and is stated in the following theorem.
\begin{Theorem}
\label{theo:pdf}
Suppose that $\mathcal{K}=\{0, \pm 1, \pm2, ..., \pm2N\}$ for some positive integer $N$, $c(k)$ is a positive semi-definite function on $\mathcal{K}$ and
\begin{equation*}
	\mathcal{S}(k)=
	\begin{cases}
		\{(-N+k+v,N-v):v=0, ..., 2N-k\} &  \quad 0\leq k \leq2N \\
		\{(N+k-v,-N+v):v=0, ..., 2N+k\} &  \quad -2N\leq k<0.
	\end{cases}
\end{equation*}
Then, $\hat{V}_{\mathcal{K},\mathcal{S}}(t)$ defined in (\ref{eq:def}) is symmetric and positive semi-definite.
\end{Theorem}
The proof of Theorem \ref{theo:pdf} is reported in the appendix \ref{sec:appprof}.

Moreover, it emerges that $\hat{V}_{\mathcal{K},\mathcal{S}}(t)$ can be rewritten as:

\begin{equation}\label{eq:pdf}
	\hat{V}^{j,j'}_N(t)=  \frac{1}{2\pi} \frac{1}{2N+1}\sum_{l=1}^{n_j}\sum_{l'=1}^{n_{j'}}\sum_{u=-N}^{N}\sum_{u'=-N}^{N}c(u-u')e^{ i u(t-t_l^j)}e^{- i u'(t-t_{l'}^{j'})} \Delta(X_l^j)\Delta(X_{l'}^{j'}),
\end{equation}
for two asset $j$ and $j'$ and $t \in (0, 2\pi)$, where $c(k)$ is still a positive semi-definite function.

\begin{Remark}\label{rem:cons}
The positive semi-definite Fourier (PDF) estimator defined above is a consistent estimator for the spot covariance, as proved in \citet{aka2023pdf}, if the function $c(k)$ is properly chosen. In particular, a Gaussian kernel is sufficient to ensure consistency, and our choice for $t \in (0, 2\pi)$ is:
	\[
	c(k) = e^{-\frac{2\pi^2k^2}{M}},\quad \text{with} \quad M \rightarrow \infty \quad \text{as} \quad N\rightarrow \infty.
	\]
To underline the presence of the localization parameter $M$, the left side equation (\ref{eq:pdf}) can be written as $\hat{V}^{j,j'}_{N,M}(t)$; insights about the sensitivity of the proposed estimator to the choice of the parameters $N$ and $M$ are reported in sections \ref{sec:choicepar} and \ref{sec:Nas}.
\end{Remark}

\begin{Remark}
By Bochner’s theorem, we know that for each positive semi-definite function $c$, there exists a bounded measure $\mu$ on $\mathbf{R}$ such that
	\[
	c(x) = \int_{\mathbf{R}}e^{2\pi i y x} \mu(dy).
	\]
Therefore, we may also rewrite the PDF estimator (\ref{eq:pdf}) using the
measure $\mu$ instead of the positive semi-definite function $c(k)$, and obtain
	\[
	\hat{V}^{j,j'}_{N, M}(t)=  \frac{1}{2\pi (2N+1)}\sum_{l=1}^{n_j}\sum_{l'=1}^{n_{j'}} \int_{\mathbf{R}} D_N(t-t_l^j+y)D_N(t-t_{l'}^{j'}+y)\mu_M(dy)\Delta(X_l^j)\Delta(X_{l'}^{j'}).
	\]
The choice of density should aligned with the choice of $c(k)$ discussed in Remark \ref{rem:cons}, for a more detailed description of the assumptions on $\mu$, see \citet{aka2023pdf}.
\end{Remark}

\section{Simulation study}\label{sec:simul}
\subsection{Simulation settings}\label{sec:sim-set}
In this section we present an extensive numerical simulated study. The aim of this study is twofold: first in section \ref{sec:choicepar} we analyze the sensitivity of the estimator to the choice of the parameters $N$ and $M$ and, with an unfeasible optimization, we find out how much the optimal choice changes in different scenarios. Secondly, in section \ref{sec:com} we evaluate the accuracy of the proposed PDF estimator and its ability to produce symmetric and positive semi-definite estimations in a comparison with two alternative estimators that are present in the literature.

To give robustness to the results of our study, in the above-mentioned exercises we consider many different simulation scenarios, focusing on both the two components of high-frequency financial data: the efficient price and the additive noise component given by market microstructure, so that the observed price $\tilde{X}$ is:
\begin{equation}
\tilde{X}^j_t = X^j_t + \eta_t^j, \qquad j=1, ..., d
\end{equation}
with $\eta$ being the noise component.

In particular, for the efficient price process we consider the following specifications:
\begin{itemize}
	\item Heston stochastic volatility model by \citet{heston1993closed};
	\item the One Factor stochastic volatility model (SVF1) used, e.g., in \citet{huang2005relative};
	\item Two Factor stochastic volatility model (SVF2) by \citet{chernov2003alternative};
	\item the Rough Heston model (RH) by \citet{el2019characteristic};
\end{itemize}
while for the additive microstructure noise we take into account the following possibilities:
\begin{itemize}
	\item no noise;
	\item noise coming from rounding;
	\item i.i.d. noise;
	\item autocorrelated noise;
	\item noise correlated with the efficient price process;
	\item heteroskedastic noise.
\end{itemize}
Since in the different cases when noise is present we analyze respectively 2, 4, 3, 3, and 3 different levels of rounding, noise intensity, noise autocorrelation, noise correlation and maximum noise intensity, in our simulated analysis we study a total of 64 different scenarios.

For simplicity of the computations, through sections \ref{sec:choicepar} - \ref{sec:com}, all the simulated analysis is conducted on the interval $[0,1]$; for that reason, and in light of Remark \ref{rem:cons}, the PDF estimator is given by:
\begin{equation}\label{eq:pdfc}
	\hat{V}^{j,j'}_{N,M}(t)=  \frac{1}{ 2N+1}\sum_{l=1}^{n_j}\sum_{l'=1}^{n_{j'}}\sum_{u=-N}^{N}\sum_{u'=-N}^{N}e^{-\frac{2\pi^2(u-u')^2}{M}}e^{ i u(t-t_l^j)}e^{- i u'(t-t_{l'}^{j'})} \Delta(X_l^j)\Delta(X_{l'}^{j'}),
\end{equation}

Where not stated otherwise, the simulations consist of $K=500$ daily trajectories, considering a trading day of length 6.5 hours, and are carried out on an equally spaced grid of width 2 seconds. To introduce asynchronicity in the data, observations are drawn from a Poisson process with intensity $\bar{\Delta}t=10$, i.e. a process that produces on average one observation every 10 seconds. Moreover, where not explicitly stated, the correlation between Brownian motions driving the efficient processes of different assets, following \citet{bibinger2019estimating}, is fixed to mimic the median estimated realized correlation of the Nasdaq components, i.e:
 \[\left\langle W^j,W^i\right\rangle =0.312, \qquad j,i=1, ..., d, \quad j \neq i .\]
In sections \ref{sec:ep} and \ref{sec:nm} we define the models used for the efficient price process and the microstructure noise.

\subsubsection{Efficient price process}\label{sec:ep}
\subsubsection*{Heston model}

The Heston stochastic volatility model by \citet{heston1993closed} is possibly the most used stochastic volatility model in the high-frequency econometric literature. It takes the form:

\begin{equation*}
	\begin{cases}
		dX^j_t &= (\mu - (\sigma^j_t)^2/2)dt + \sigma^j_t dW^j_t\\
		d(\sigma^j_t)^2 &= \gamma(\theta-(\sigma^j_t)^2)dt + \nu\sigma^j_t dZ^j_t,
	\end{cases}
\end{equation*}
with $\left\langle W^j,Z^j\right\rangle =\lambda$ to account for the leverage effect, and $\left\langle W^j,Z^i\right\rangle =0$ for $i\not=j$.

The parameters are set to be:
\[
(\mu,\gamma, \theta, \nu, \lambda) = (0.05/252, 5/252, 0.1, 0.5/252, -0.5),
\]
that is the same choice made by \citet{zu2014estimating}, \citet{mancino2015fourier} and \citet{figueroa2022kernel}.

\subsubsection*{Factor volatility models}
Factor volatility models have been long used in the literature, see for example \citet{huang2005relative}. First we consider the One Factor Stochastic Volatility model (SV1F) of the form:

\begin{equation*}
	\begin{cases}
		dX^j_t &= \mu dt+ \sigma^j_t dW^j_t\\
		\sigma^j_t &=\textrm{e}^{\beta_0+\beta_1 \tau^j_t} \qquad \qquad\\
		d\tau^j_t&= \alpha \tau^j_t+ dZ^j_t
	\end{cases}
\end{equation*}
for $j=1,...,d$, with $\left\langle W^j,Z^j\right\rangle =\lambda$ , and $\left\langle Z^j,Z^{j'}\right\rangle = 0$ for $j\neq j'$.
The simulation is conducted using as parameters:

\[
(\mu, \beta_1, \alpha, \beta_0, \lambda) = (0.03, 0.125, -0.025, \beta_1/(2\alpha), -0.3)
\]
that are the parameters used also in \citet{zu2014estimating}, \citet{mancino2015fourier} and \citet{figueroa2022kernel} and \citet{mancino2022asymptotic}.

Second, we consider the Two Factors Stochastic Volatility model (SV2F), proposed by \citet{chernov2003alternative}, and able to reproduce higher values of volatility of volatility. It has the form:
\begin{equation*}
	\begin{cases}
		dX^j_t &= \mu dt+ \text{s-exp}[\beta_0+\beta_1 \tau^{j,1}_t +\beta_2 \tau^{j,2}_t]dW^j_t\\
		d\tau^{j,1}_t&= \alpha_1 \tau^{j,1}_t + dZ^{j,1}_t\\
		d\tau^{j,2}_t&= \alpha_2 \tau^{j,2}_t + (1+\beta_v\tau^{j,2}_t) dZ^{j,2}_t
	\end{cases}
\end{equation*}
with
\begin{equation*}
\text{s-exp(x)}=
\Bigg\{ \begin{array}{ll}
	
		\exp(x), & \text{if} \quad x\leq x_0 = \log(1.5)\\
		\frac{e^{x_0}}{\sqrt{x_0}}\sqrt{x_0-x_0^2+x^2}, & \text{otherwise}
\end{array}
\end{equation*}
for $j=1,...,d$, with $\left\langle W^j,Z^{j,1}\right\rangle = \left\langle W^j,Z^{j,2}\right\rangle=\lambda$ , and $\left\langle Z^{j,i},Z^{j',i'}\right\rangle = 0$ for $j\neq j'$ or $i\neq i'$, $i,i'=1,2$.
For the parameters our choice is to use:

\[
(\mu, \beta_0, \beta_1, \beta_2, \beta_v, \alpha_1, \alpha_2, \lambda) =(0.03, -1.1, 0.04, 0.3, -0.003, -0.6, 0.25).
\]

\subsubsection*{Rough volatility}
A new strand of financial econometric literature has grown considering dynamics of the volatility process that are not driven by a standard Brownian motion, but instead are driven by a fractional Brownian motion, with Hurst index $H < 0.5$, see \citet{alos2007short}, \citet{gatheral2018volatility} and \citet{euch2018perfect}. The proposed PDF estimator is consistent even in presence of rough volatility, see \citet{aka2023pdf}.

Rough volatility may also be modeled through stochastic Volterra equation, as in the Rough Heston model studied by \citet{el2019characteristic} and that we intend to use in this analysis:
\begin{equation*}
	\begin{cases}
		X^j_t &= X^j_0+\int_0^t X^j_t \sigma_s^j dW^j_s\\
		(\sigma^j_t)^2&= (\sigma^j_0)^2 +\int_0^t K(t-s)\left( (\theta-\gamma (\sigma^j_s)^2)ds + \nu \sigma^j_s dZ^j_s\right)
	\end{cases}
\end{equation*}
with $\left\langle W^j,Z^j\right\rangle =\lambda$ and $K(t)=Ct^{H-\frac{1}{2}}$ for a Hurst index $H \in \left( 0, \frac{1}{2}\right)$ and a constant $C$.
In order to simulate the rough Heston model we apply the discrete-time Euler-type scheme studied in \citet{richard2021discrete}, referring in particular to equation (8) thereof. The parameters of the model are set to ensure that in the exercise the simulated volatility process does not exhibits negative values, and in particular they take values:
\[
(\theta, \gamma, \nu, \lambda, H) = (0.2, 0.3, 0.2, -0.7, 0.1),
\]
where the choice for the Hurst parameters is driven by the empirical evidence present in the literature; see, e.g., \citet{gatheral2018volatility}.

\subsubsection{Market microstructure noise specifications}\label{sec:nm}
It is a know fact (see, e.g., \citet{bandi2008microstructure}) that high-frequency data are contaminated by the so called market microstructure noise. In particular, a common representation of the observed price considers the sum of the efficient price and the noise component. The origin of noise is linked to specific characteristics of the  microstructure of financial markets, such as bid-ask spread, rounding, strategic trading (see, e.g. \citet{hasbrouck2007empirical}), and several models specification for the noise have been proposed in the literature of high-frequency financial econometrics.

\subsubsection*{Noise coming form rounding}
In presence of rounding the observed price process has the following form:
\[
\tilde{X}^j_t=\log([\exp(X^j_t)/r]r),
\]
where $X$ denotes the efficient price process. We consider two levels of rounding, corresponding to $r=1$ or $5$ cents, which are the most used in financial markets.

\subsubsection*{Noise i.i.d.}
The most widely used characterization of noise is to consider it an i.i.d. additive component, with mean equal to zero and a given variance:

\begin{equation}\label{eq:noise}
	\tilde{X}^j_t=X^j_t+\eta^j_t,
\end{equation}
\[\eta^j \sim i.i.d. \quad E[\eta^j]=0, \quad E[(\eta^j)^2]= var(\Delta X^j_{10sec})\sigma^2_{\eta},
\]
where $X^j_{10sec}$ denotes the regularly-spaced series obtained sampling the simulated series every 10 seconds.

Here we specify a Gaussian distribution for the noise, and we set its variance proportional to the average time between two consecutive observations $\bar{\Delta}t$, to avoid that the noise is wiped out by subsampling effect. We consider four values for the variance of the noise: $\sigma^2_\eta=1,1.5,2,2.5$.

\subsubsection*{Autocorrelated noise.}
In this case, autocorrelation is introduced in the noise component and the noise, while maintaining the additive form of equation (\ref{eq:noise}), is modeled through an Ornstein-Uhlenbeck process defined as:
\[
d\eta^j_t=-\theta_{\eta}\eta^j_tdt+\sigma_{\eta}dE^j_t,
\]
where $E$ is a standard Brownian motion independent of $W$. Three different levels of autocorrelation are considered, using $\theta_\eta=0.2, 0.3, 0.4$. $\sigma^2_\eta$ is set to obtain the same level of variance as the second case in the previous section.

\subsubsection*{Noise correlated with the efficient price process}
A natural generalization of the previous specification of the noise is to allow for correlation between the Brownian motion that drives the noise component and the one behind the efficient price dynamic:
\[
d\eta^j_t=-\theta_{\eta}\eta^j_tdt+\sigma_{\eta}dE^j_t,\qquad \left\langle W^j,E^j\right\rangle =\rho_{\eta}
\]
where $\rho_\eta$ is a constant. In light of the empirical evidence present in the literature (see, e.g. \citet{hansen2006realized}), we impose $\rho_\eta$ to be negative, with three considered values: -0.1, -0.3, -0.5, and we consider a level of noise equal to the one obtained in the i.i.d. case with noise to signal ratio equal to 2, and absence of autocorrelation in the noise process is assumed.

\subsubsection*{Heteroskedastic noise}
In the last formulation for MMN we allow for the variance of noise to be time changing over the interval. In particular, to reproduce the stylized fact observed by \citet{kalnina2008estimating} that the noise is higher at the beginning and at the end of the trading day, following \citet{bu2022nonparametric}, we use the function:

\[
\sigma_{\eta}(t)=\bar{\sigma}_{\eta}\left(\frac{1}{2}(\cos(2\pi t)+1)0.9+0.1\right)
\]
choosing $\bar{\sigma}_\eta=3, 3.5, 4$. For the noise we maintain for this case the Ornstein-Uhlenbeck formulation stated above, considering also the presence of both autocorrelation and dependence between noise and efficient price process, with $\theta=0.3$ and $\rho_n=-0.3$.

This last formulation allow for a quite general form of microstructure noise, that includes the stylized facts observed in the literature. In a last, unreported, exercise, we add also a rounding of 1 cent to this simulation scheme, without significant changes in the results.

\subsection{Selection of {\bf frequencies} $N$ and $M$}\label{sec:choicepar}

In this section we want to evaluate the sensitivity of the estimator to the choice of the frequencies $N$ and $M$ appearing in the definition of the PDF estimator. The performance of the estimator for each couple of parameters is evaluated in order to search for the optimal parameters over the entire time interval, across the $K$ simulated independent trajectories considered in our unfeasible optimization exercise. In all the following analysis the spot variance trajectory is reconstructed on a regular grid of width 20 minutes.

In this optimization study we let the values of the cutting frequency $N$ to be dependent on $n$, while the localization parameter $M$ takes values that depend on $N$ itself, building a grid of couples $(N,M)$, with values:
\[
N=\frac{n^\alpha}{2}, \qquad \alpha = 1, \frac{5}{6}, \frac{3}{4}, \frac{2}{3}, \frac{1}{2}, \frac{1}{3},
\]
\[
M=N^\beta, \qquad \beta= \frac{5}{6}, \frac{3}{4}, \frac{2}{3}, \frac{1}{2}, \frac{4}{9},
\]
where the choice for the possible values of $N$ relies on the known fact that, as for all the other Fourier-type estimators, in order to avoid aliasing effects, the cutting frequency must be smaller than the Nyquist frequency, i.e., {\bf $N \leq n/2$}.

For each scenario, in a setting with $d=2$, on the grid of values described above, we look at the error made in estimating the variance $\hat{V}^{1,1}$ and the covariance $\hat{V}^{1,2}$, using in particular the integrated error:
\[
MISE_j=(K)^{-1}\sum_{k=1}^K\int_0^1(\hat{V}^{1,j}_k-{V}^{1,j}_k)^2dt, \qquad j=1,2
\]
and choosing as optimal the pair that minimize
\[
0.1\cdot MISE_1 + 0.9\cdot MISE_2,
\]
where a higher weight is given to the estimation of the covariance, being the dominant component of a generic variance-covariance matrix.

Table \ref{tab:opt} shows the optimal couple of $N$ and $M$ for each scenario. The first thing to notice is that the optimal value for $N$ seems to be pretty stable across the different model for the efficient price process, and it is always smaller than the Nyquist frequency $n/2$, even in absence of noise. When the data are more heavily affected by noise, the optimal $N$ is smaller, coherently with what is known for the other Fourier-type estimators of both integrated and spot volatility, see, e.g., \citet{mancino2017fourier}. Moreover, the choice of $N$, while being sensitive to the intensity of noise, does not seems to be dependent of its correlation structure. For what concerns the parameter $M$, it seems to have almost the same optimum in all the scenarios considered in this exercise. Moreover, for the four models for the efficient price, the difference between $M=N^{4/9}$ and $M=N^{1/2}$ seems to be negligible, as shown in table \ref{tab:optex}, in which $MISE_2$ on the defined grid is reported for selected scenarios. From table \ref{tab:optex} it is also clear that, overall, the estimator is much more sensible to the choice of $N$ than to the choice of $M$, and changes in the former cause higher shifts in the estimation error; even if, for sake of simplicity, table \ref{tab:optex} shows the figures only for the Heston and the SVF2 models, the behavior for the remaining models is analogous. It is important stress that, in line with what observed also for the original spot Fourier estimator in \citet{mancino2022asymptotic}, the removal of the noise is carried out with an appropriate choice of the cutting frequency $N$. In fact, $N$ controls the number of Fourier coefficients of the asset returns to be considered in the convolution to obtain the Fourier coefficient of the covariance process. It is well know that by reducing the number of frequencies $N$, the noise is filtered out (see \citet{mancino2015fourier}.

\begin{table}[htbp!]		
	\begin{center}		
		\begin{tabular}{C|C|C|C|C}
			\hline\hline
			 & Heston & SVF1 & SVF2 & RH \\
			\hline\hline
			/ & \multicolumn{4}{c}{No noise}\\
			\hline
			/  & 3/4, 4/9  & 3/4, 4/9 & 3/4, 1/2 & 3/4, 1/2 \\
			\hline
			r & \multicolumn{4}{c}{Noise from rounding}\\
			\hline
			0.01 & 3/4, 4/9 & 3/4, 4/9  & 3/4, 1/2  & 3/4, 1/2\\
			0.05 & 3/4, 4/9 & 3/4, 4/9 & 3/4, 1/2 & 3/4, 1/2 \\
			\hline
			$\sigma_{\eta}$ & \multicolumn{4}{c}{I.i.d. noise}\\
			\hline
			1 & 3/4, 4/9 & 3/4, 4/9  & 3/4, 1/2 & 3/4, 1/2 \\
			1.5 & 3/4, 4/9 & 3/4, 4/9 & 3/4, 1/2 &  3/4, 1/2\\
			2 & 2/3, 4/9 & 2/3, 4/9 & 2/3, 1/2 & 2/3, 1/2 \\
			2.5 & 2/3, 4/9 & 2/3, 4/9 & 2/3, 1/2 & 2/3, 1/2\\
			\hline
			$\theta$ & \multicolumn{4}{c}{Autocorrelated noise}\\
			\hline
			0.2 & 2/3, 4/9 & 2/3, 4/9  & 2/3, 1/2  & 2/3, 1/2 \\
			0.3 & 2/3, 4/9 & 2/3, 4/9 & 2/3, 1/2 & 2/3, 1/2  \\
			0.4 & 2/3, 4/9 & 2/3, 4/9 & 2/3, 1/2  & 2/3, 1/2\\
			\hline
			$\rho_{\eta}$ & \multicolumn{4}{c}{Noise correlated with $p$}\\
			\hline
			-0.1 & 2/3, 4/9 & 2/3, 4/9 & 2/3, 1/2 & 2/3, 1/2 \\
			-0.3 & 2/3, 4/9 & 2/3, 4/9 & 2/3, 1/2 & 2/3, 1/2 \\
			-05  & 2/3, 4/9 & 2/3, 4/9 & 2/3, 1/2 & 2/3, 1/2 \\
			\hline
			$\bar{\sigma}_{\eta}$ & \multicolumn{4}{c}{Heteroskedastic noise}\\
			\hline
			3 & 2/3, 4/9 & 2/3, 4/9 & 2/3, 1/2 & 2/3, 1/2 \\
			3.5 & 2/3, 4/9 & 2/3, 4/9 & 2/3, 1/2 & 2/3, 1/2 \\
			4 & 2/3, 4/9 & 2/3, 4/9 & 2/3, 1/2 & 2/3, 1/2 \\
			\hline
		\end{tabular}
		\caption{Optimal couple of $\alpha$, $\beta$ in the considered grid across the different models for volatility and microstructure noise.}\label{tab:opt}
	\end{center}
\end{table}

\begin{table}[htbp!]
		\scriptsize
		\hspace{-2.0cm}
		\begin{tabular}{c|ccccc}
			\hline\hline
			\multicolumn{6}{c}{Heston - No noise}\\
			\hline\hline
			$\alpha$/$\beta$ & 5/6 & 3/4 & 2/3 & 1/2 & 4/9\\
			\hline
			1 & 1.0386e-03 & 1.0204e-03  & 1.0069e-03 &  9.9026e-04   & 9.8691e-04 \\
			5/6 & 2.3748e-04 &  2.0522e-04 &  1.7974e-04 &  1.4534e-04  & 1.3932e-04 \\
			3/4 & 2.3898e-04 &  1.9593e-04 &  1.6083e-04 &  1.1666e-04  & 1.1096e-04 \\
			2/3 & 3.3382e-04 &  2.8090e-04 & 2.3916e-04  & 1.9747e-04  & 1.9399e-04 \\
			1/2 & 7.4551e-04  & 7.0530e-04 &  6.8504e-04 &  6.7519e-04  & 6.7487e-04 \\
			1/3 & 2.0294e-03  & 2.0282e-03  & 2.0278e-03 &  2.0277e-03  & 2.0277e-03 \\
			\hline
		\end{tabular}\hspace{0.5cm}
			\begin{tabular}{c|ccccc}
		\hline\hline
		\multicolumn{6}{c}{SVF2 - No noise}\\
		\hline\hline
		$\alpha$/$\beta$ & 5/6 & 3/4 & 2/3 & 1/2 & 4/9\\
		\hline
		1 & 1.4352e-03  & 1.4189e-03 &  1.4101e-03  & 1.4065e-03 & 1.4081e-03\\
		5/6 & 4.0236e-04  & 3.7224e-04 &  3.5338e-04  & 3.3913e-04 & 3.4125e-04\\
		3/4 & 3.6879e-04  & 3.2276e-04 &  2.9026e-04  & 2.6256e-04 &  2.6423e-04\\
		2/3 & 4.2664e-04  & 3.7789e-04 &  3.4306e-04  & 3.2132e-04 &  3.2378e-04\\
		1/2 & 8.8461e-04  & 8.5025e-04 &  8.3790e-04  & 8.4012e-04 &  8.4159e-04\\
		1/3 & 2.9639e-03  & 2.9649e-03 &  2.9659e-03  & 2.9668e-03 &  2.9669e-03\\
		\hline
	\end{tabular}\vspace{0.3cm}

	\hspace{-2.0cm}
	\begin{tabular}{c|ccccc}
	\hline\hline
	\multicolumn{6}{c}{Heston - I.i.d. noise, $\sigma_{\eta}=2.5$}\\
	\hline\hline
	$\alpha$/$\beta$ & 5/6 & 3/4 & 2/3 & 1/2 & 4/9\\
	\hline
	1 & 6.6533e-03 &  5.2534e-03 &  4.2323e-03 &  2.9124e-03 &  2.6191e-03\\
	5/6 & 1.7481e-03 &  1.3619e-03 &  1.0611e-03 &  6.6181e-04 &  5.9198e-04\\
	3/4 & 7.3901e-04 &  6.0224e-04 &  4.9336e-04 &  3.5747e-04 &  3.3973e-04\\
	2/3 & 4.7485e-04 &  3.9982e-04 &  3.4147e-04 &  2.8342e-04 &  2.7856e-04\\
	1/2 & 7.7739e-04 &  7.3628e-04 &  7.1565e-04 &  7.0578e-04 &  7.0548e-04\\
	1/3 & 2.0670e-03 &  2.0658e-03 &  2.0654e-03 &  2.0653e-03 &  2.0653e-03\\
	\hline
\end{tabular}\hspace{0.5cm}
\begin{tabular}{c|ccccc}
	\hline\hline
	\multicolumn{6}{c}{SVF2 - I.i.d. noise, $\sigma_{\eta}=2.5$}\\
	\hline\hline
	$\alpha$/$\beta$ & 5/6 & 3/4 & 2/3 & 1/2 & 4/9\\
	\hline
    1 & 9.3093e-03 &  7.1797e-03 &  5.5577e-03 &  3.4186e-03 &  2.9621e-03\\
	5/6 & 2.8428e-03 &  2.2971e-03 &  1.8831e-03 &  1.3697e-03 & 1.2876e-03\\
	3/4 & 1.2835e-03 &  1.0975e-03 &  9.5587e-04 &  7.9035e-04 &  7.7287e-04\\
	2/3 & 6.1650e-04 &  5.3221e-04 &  4.6916e-04 &  4.2024e-04 &  4.2082e-04\\
	1/2 & 9.1933e-04 &  8.8559e-04 &  8.7380e-04 &  8.7665e-04 &  8.7821e-04\\
	1/3 & 3.0359e-03 &  3.0369e-03 &  3.0380e-03 &  3.0390e-03 &  3.0391e-03\\
	\hline
\end{tabular}
	\caption{Error in estimating covariance over the considered grid, for selected scenarios.}\label{tab:optex}
\end{table}

\begin{Remark}
	While in this exercise and in the following analysis, focusing on obtaining the best level of performance for the whole variance-covariance matrix for a single couple of parameters $N$ and $M$, a higher weight is given to covariance estimation, it is worth noting that, when dealing with estimating variance alone (i.e. in an univariate setting) the optimal choice is usually different from the one obtained for covariance, for the differences in the fluctuations of the two quantities. In particular, as table \ref{tab:optexvar} shows for the case of the Rough Heston model, in absence of noise the best choice is obtained for higher values of $N$ and $M$ (in particular setting $\alpha=1$), while in presence of noise the differences are reduced, leading to a best choice of $\alpha=2/3$ and $\beta=5/6$. The values for $N$ are in line with the one observed in similar experiments for the traditional Fourier estimator of spot volatility, see, e.g., \citet{mancino2015fourier}, while the impact of $M$ is confirmed to be less pronounced also in this case. More insights on the differences in optimizing variance and covariance in presence of asynchronicity are given in the next section.
	
	\begin{table}[htbp!]
		\scriptsize
		\hspace{-2.3cm}
		\begin{tabular}{c|ccccc}
			\hline\hline
			\multicolumn{6}{c}{Rough Heston - No noise}\\
			\hline\hline
			$\alpha$/$\beta$ & 5/6 & 3/4 & 2/3 & 1/2 & 4/9\\
			\hline
			1 & 6.4138e-03 &	6.7283e-03	& 7.1840e-03 &	8.0398e-03 &	8.8810e-03\\
			5/6 & 7.5422e-03 &	7.8732e-0.3 &	8.2934e-03 &	9.3735e-03 &	9.8061e-03 \\
			3/4 & 8.4182e-03 &	8.7230e-03 & 9.1001e-03 &	1.0076e-02 &	1.0431e-02 \\
			2/3 & 9.5777e-03 &  9.7728e-03 & 1.0058e-03 &	1.0887e-02 &	1.1144e-02 \\
			1/2 & 1.3381e-02 &	1.3525e-02	& 1.3756e-02 &	1.4132e-02 &	1.4190e-02 \\
			1/3 & 2.5303e-02 &	2.5353e-02 & 2.5384e-02 & 2.5407e-02 &	2.5409e-02\\
			\hline
		\end{tabular}\hspace{0.4cm}
		\begin{tabular}{c|ccccc}
			\hline\hline
			\multicolumn{6}{c}{Rough Heston -  noise}\\
			\hline\hline
			$\alpha$/$\beta$ & 5/6 & 3/4 & 2/3 & 1/2 & 4/9\\
			\hline
			1 & 4.6703e+00 &	4.6520e+00 &	4.6384e+00 & 	4.6204e+00 &	4.6161e+00\\
			5/6 & 4.1665e-01 &	4.1473e-01 &	4.1331e-01 &	0.41143-01 &	4.1117e-01 \\
			3/4 & 6.2516e-02 &	6.2070e-02 &	6.1793e-02 &	6.1843e-02 &	6.2042e02\\
			2/3 & 1.3675e-02 &	1.3813e-02 &	1.4038e-02 &	1.4408e-02 & 1.4464e-02\\
			1/2 & 1.4980e-02 &	1.4994e-02 &	1.5127e-02 &	1.5778e-02 &	1.6012e-02\\
			1/3 & 2.6021e-02 &	2.6071e-02 &	2.6102e-02 &	2.6125e-02 &	2.6127e-02\\
			\hline
		\end{tabular}
	\caption{Error in estimating variance over the considered grid, for the Rough Heston model.}\label{tab:optexvar}
\end{table}
\end{Remark}

\subsubsection{The impact of introducing asynchronicity in covariance estimation}\label{sec:Nas}

In this section we study more in detail how much the effect of asynchronicity in the data affects the optimal choice of the cutting frequency $N$, since introduction of asynchronicity might influence the efficiency of spot variance estimation, as shown in \citet{park2016estimating}, and may lead to a different optimal value for the parameters of a Fourier-type estimator, as shown in \citet{mancino2017fourier} for the case of integrated variance. While in this section we compare the synchronous and the asynchronous case to evaluate the impact on the choice of $N$, an additional analysis on the influence of sampling schemes on the error of estimation is presented in section \ref{sec:sim-noise}.

We work here in a simplified environment along the lines of \citet{mancino2017fourier}, and we maintain in this section a fixed $M=N^{4/9}$, in light of the results of the previous section. In particular we consider two asset, whose prices $p^1$ and $p^2$ are equal to two standard Brownian Motions $W^1$ and $W^2$, with $\left\langle W^1,W^2\right\rangle = \rho$. To induce asynchronicity in the observations here we simply introduce two distinct vectors of observation times:
\[t^1_i= i/n \quad \text{for} \quad i=0,...,n \quad \text{and}\]
\[t^2_i = i/n + 0.5 \quad \text{for} \quad i=1,...,n-1,\]
with $t^2_0=0$ and $t^2_n = 1$. We set $n=500$ and we consider a set of 10000 simulated price paths, on which we estimate the covariance between $W^1$ and $W^2$ considering two scenarios for the sampling scheme: the first one in which the two assets are observed synchronously on to the time grid $t^1$, and the second in which the the two assets are observed respectively on $t^1$ and $t^2$.
$\hat{V}^{1,2}_N(t)$ is computed for $N=0,...,n$ and for $t=0.5$ (i.e. in the middle of the time window), to evaluate the performance of the estimators while varying the parameter $N$. The analysis is carried out considering multiple values of correlation between $W^1$ and $W^2$, namely $\rho =\left\lbrace  0.2, \pm0.3, \pm0.5, \pm0.7, \pm1 \right\rbrace $, so that, under this point of view, we expand the range possibilities with respect to the previous section, where a single value for the correlation between assets was taken into account.

\begin{figure}[htpb!]
	\centering
	\includegraphics[width=0.43\linewidth]{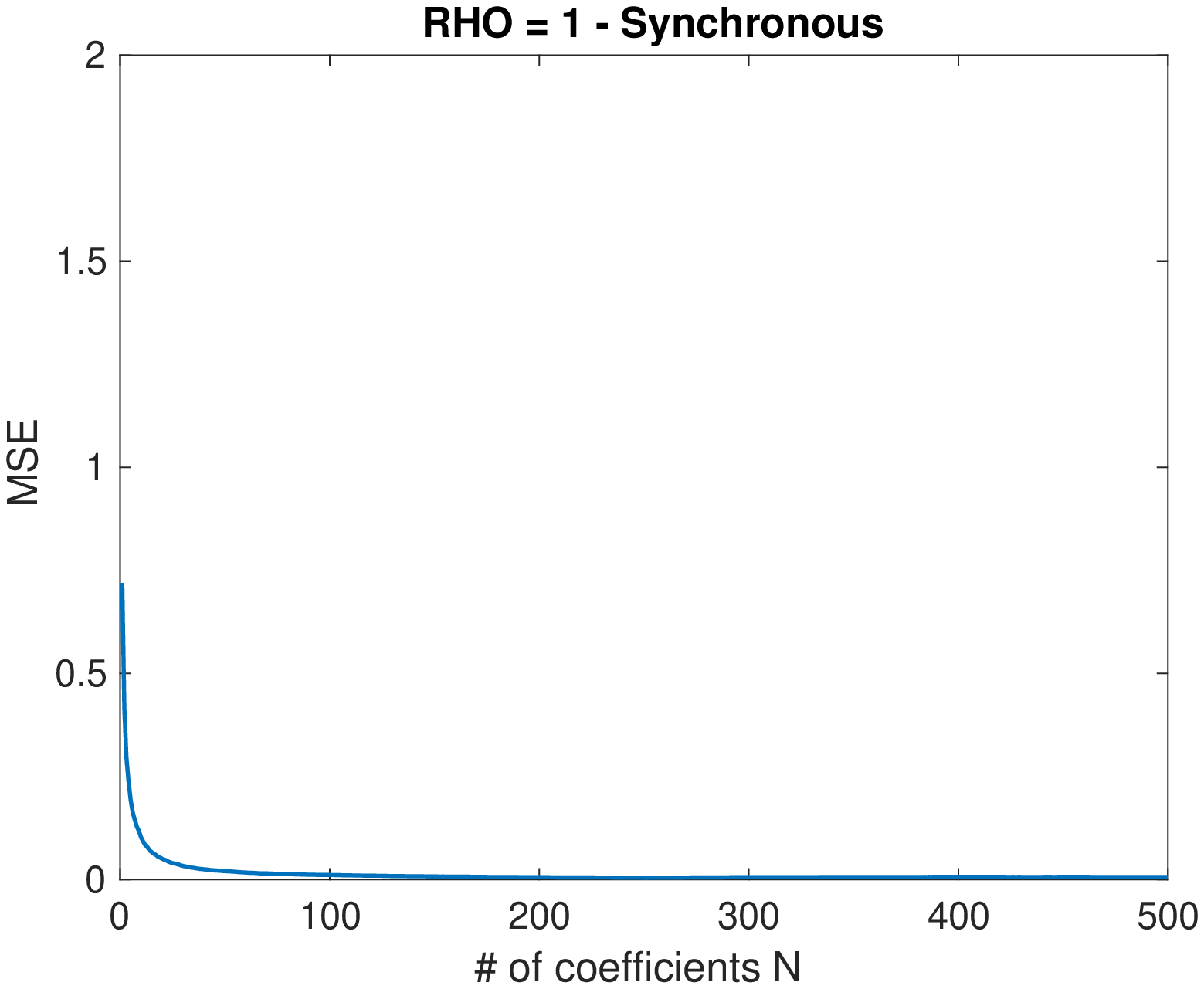}
	\includegraphics[width=0.43\linewidth]{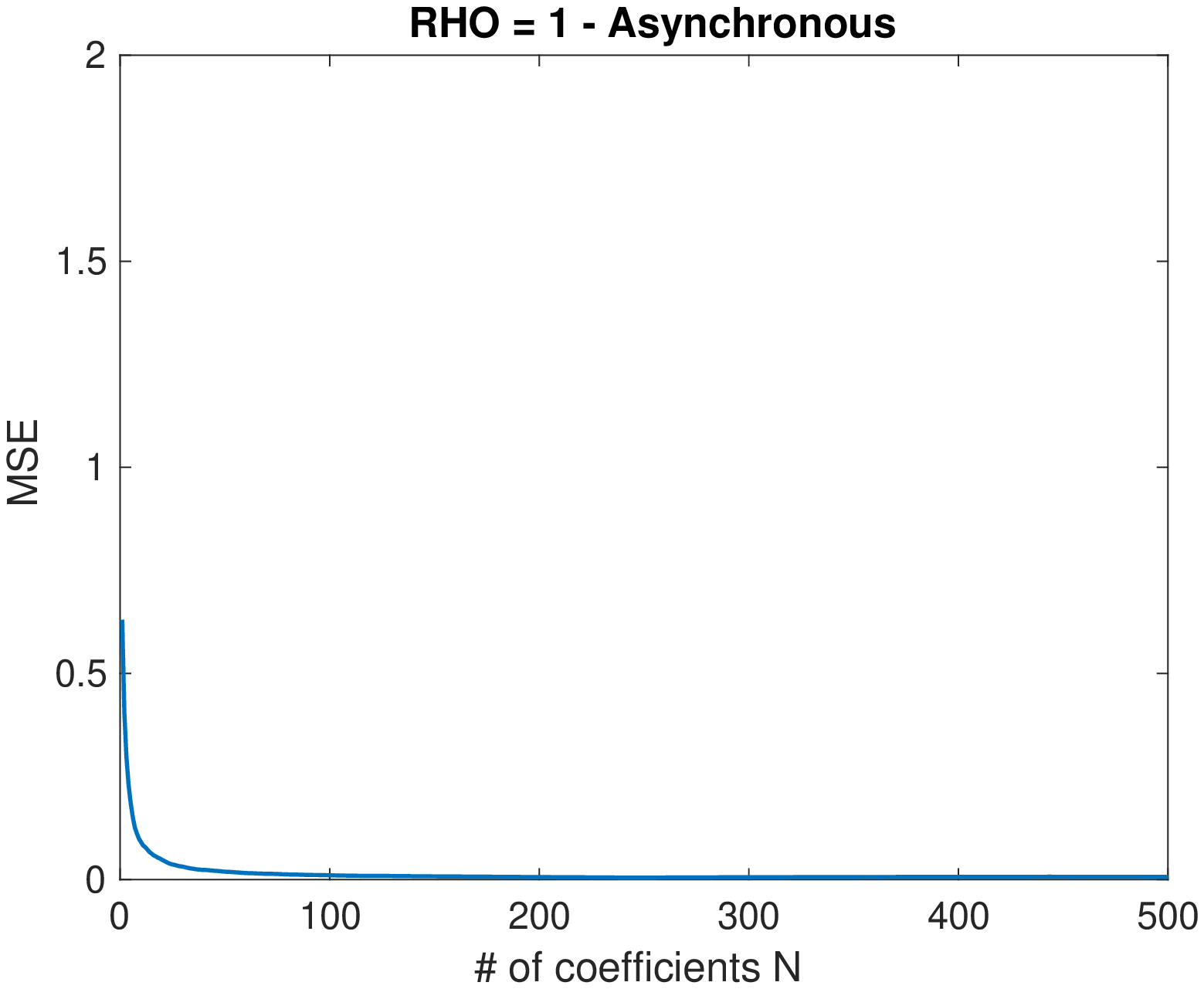}\\
	\includegraphics[width=0.43\linewidth]{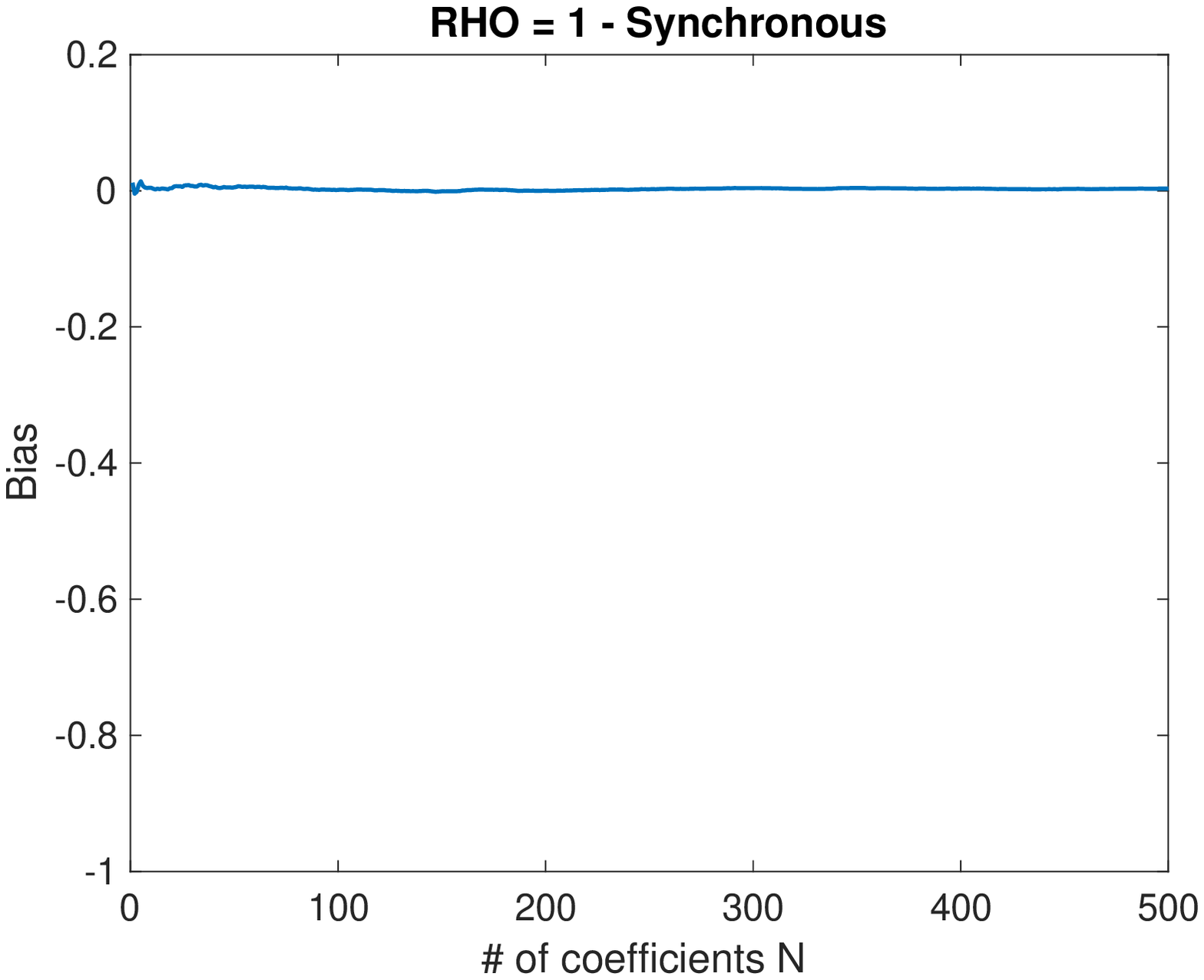}
	\includegraphics[width=0.43\linewidth]{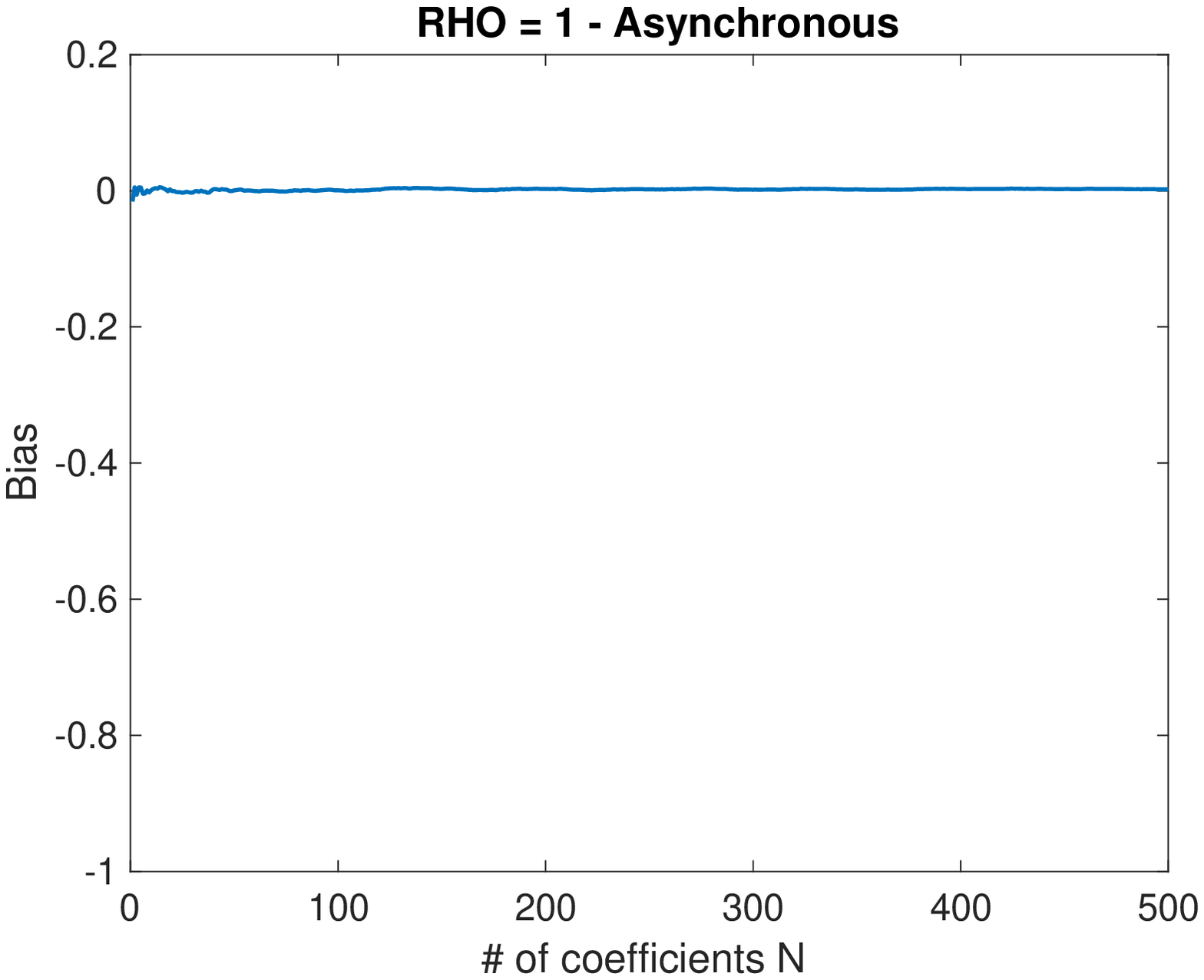}
	\caption{Mean square error (first row) and bias (second row) for $\rho=1$.}
	\label{fig:as1}
\end{figure}

\begin{figure}[htpb!]
	\centering
	\includegraphics[width=0.43\linewidth]{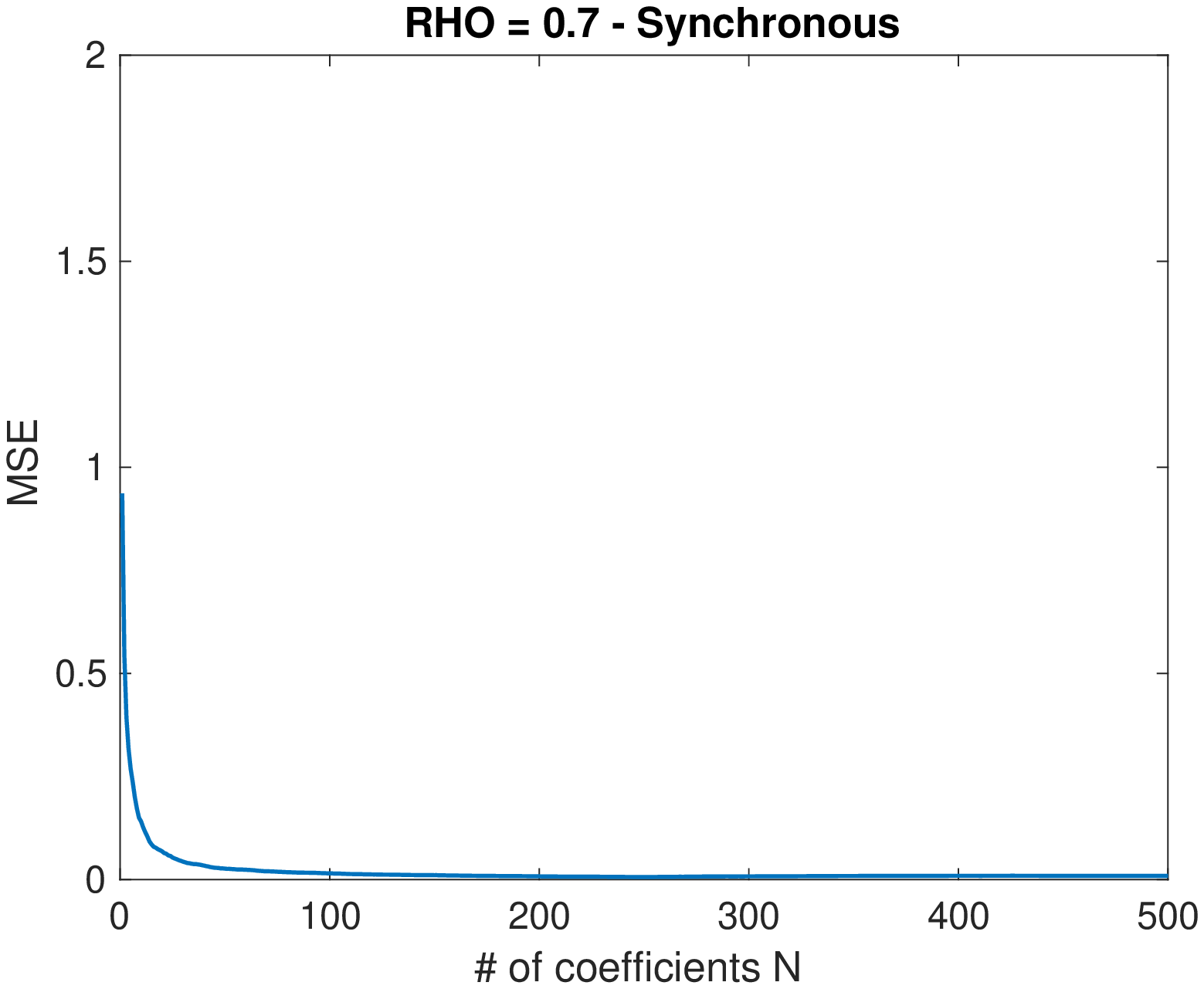}
	\includegraphics[width=0.43\linewidth]{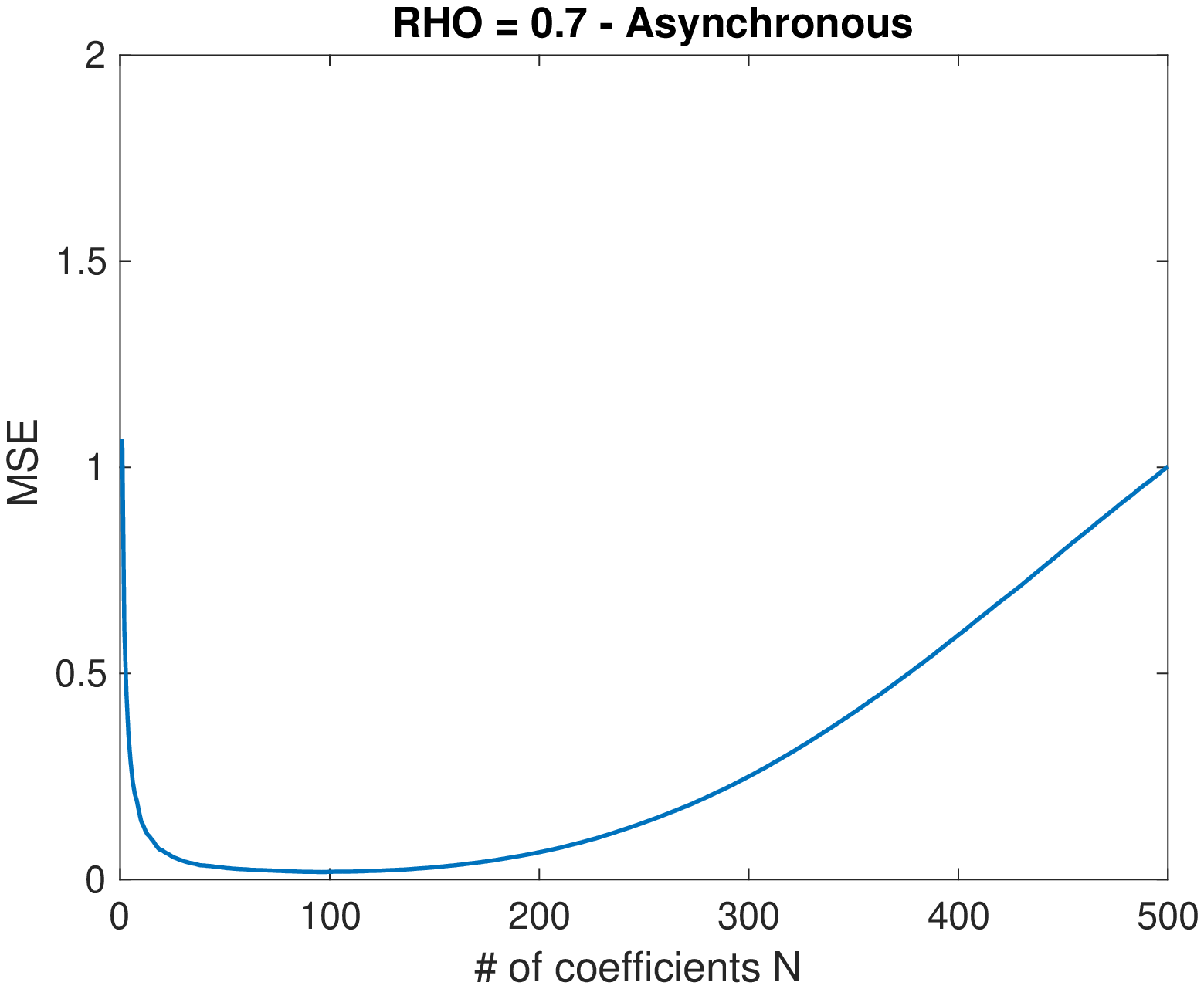}\\
	\includegraphics[width=0.43\linewidth]{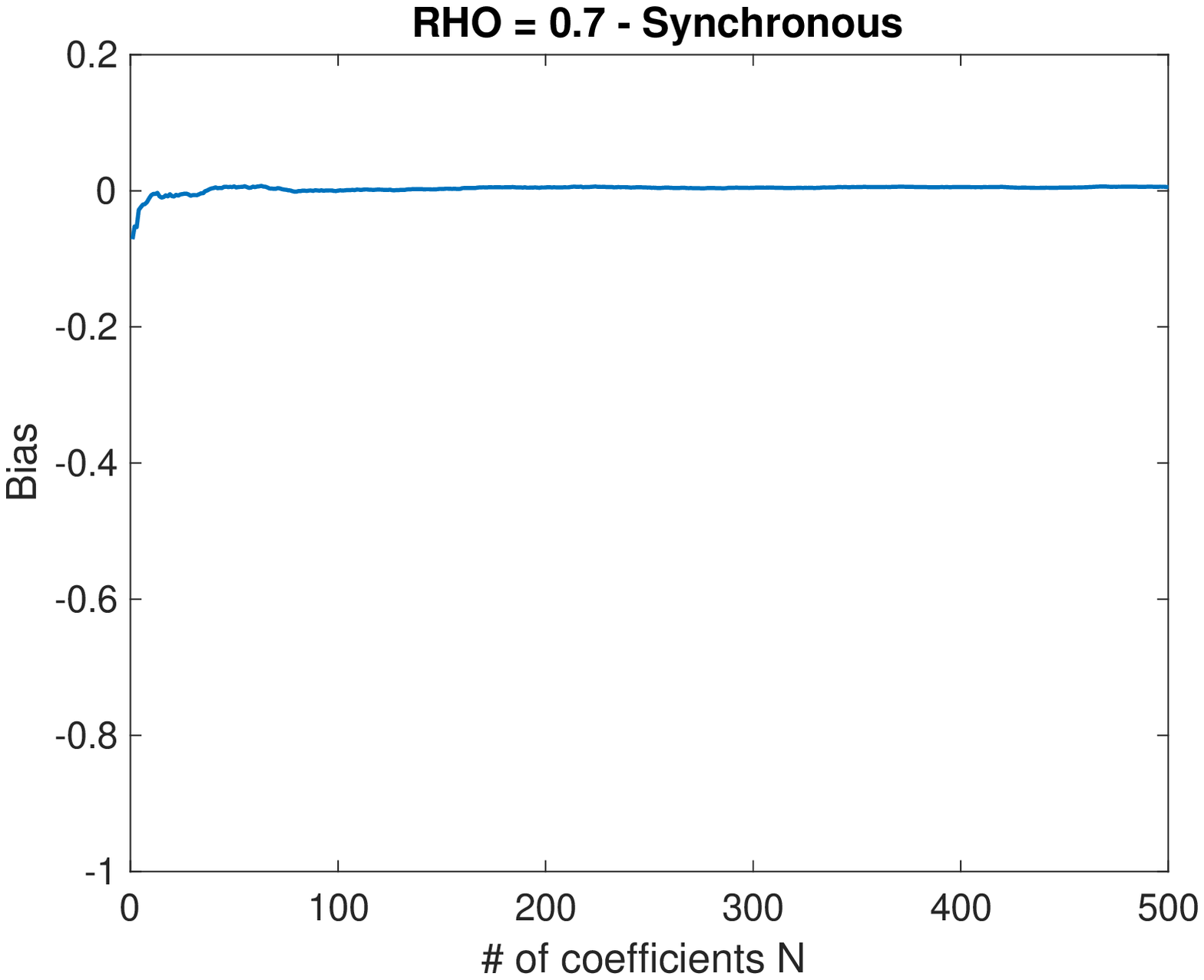}
	\includegraphics[width=0.43\linewidth]{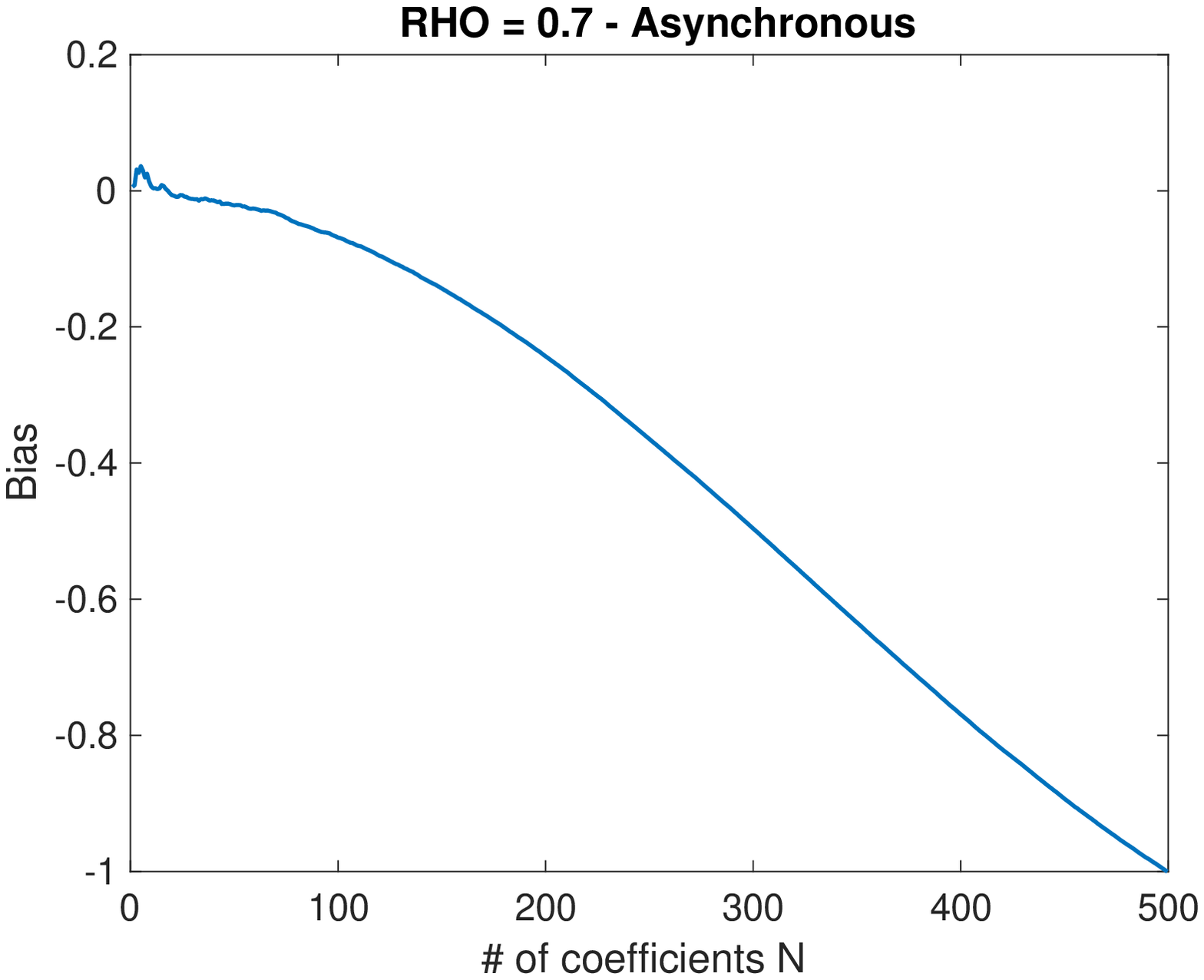}
	\caption{Mean square error (first row) and bias (second row) for $\rho=0.7$.}
	\label{fig:as07}
\end{figure}

\begin{figure}[htpb!]
	\centering
	\includegraphics[width=0.43\linewidth]{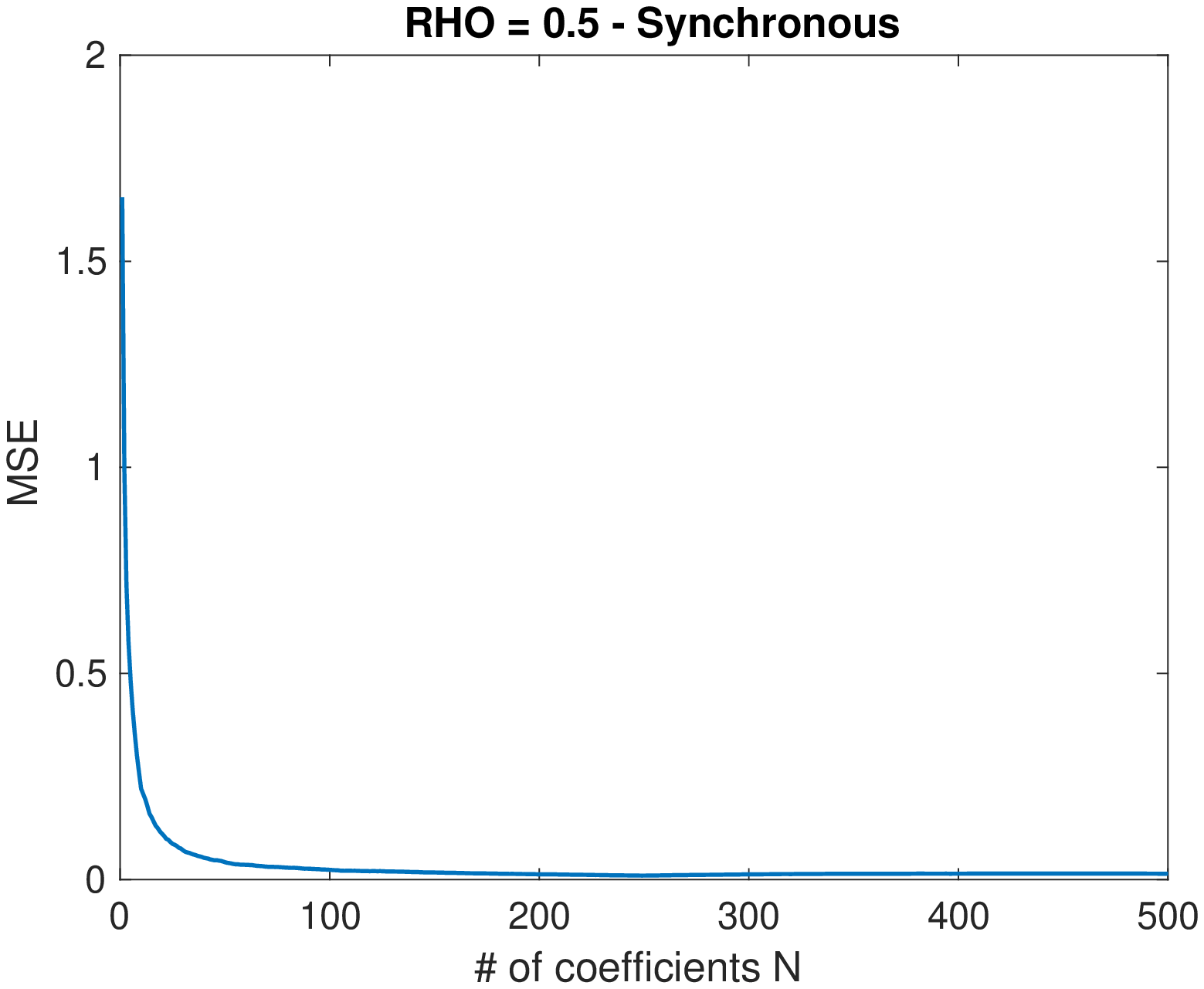}
	\includegraphics[width=0.43\linewidth]{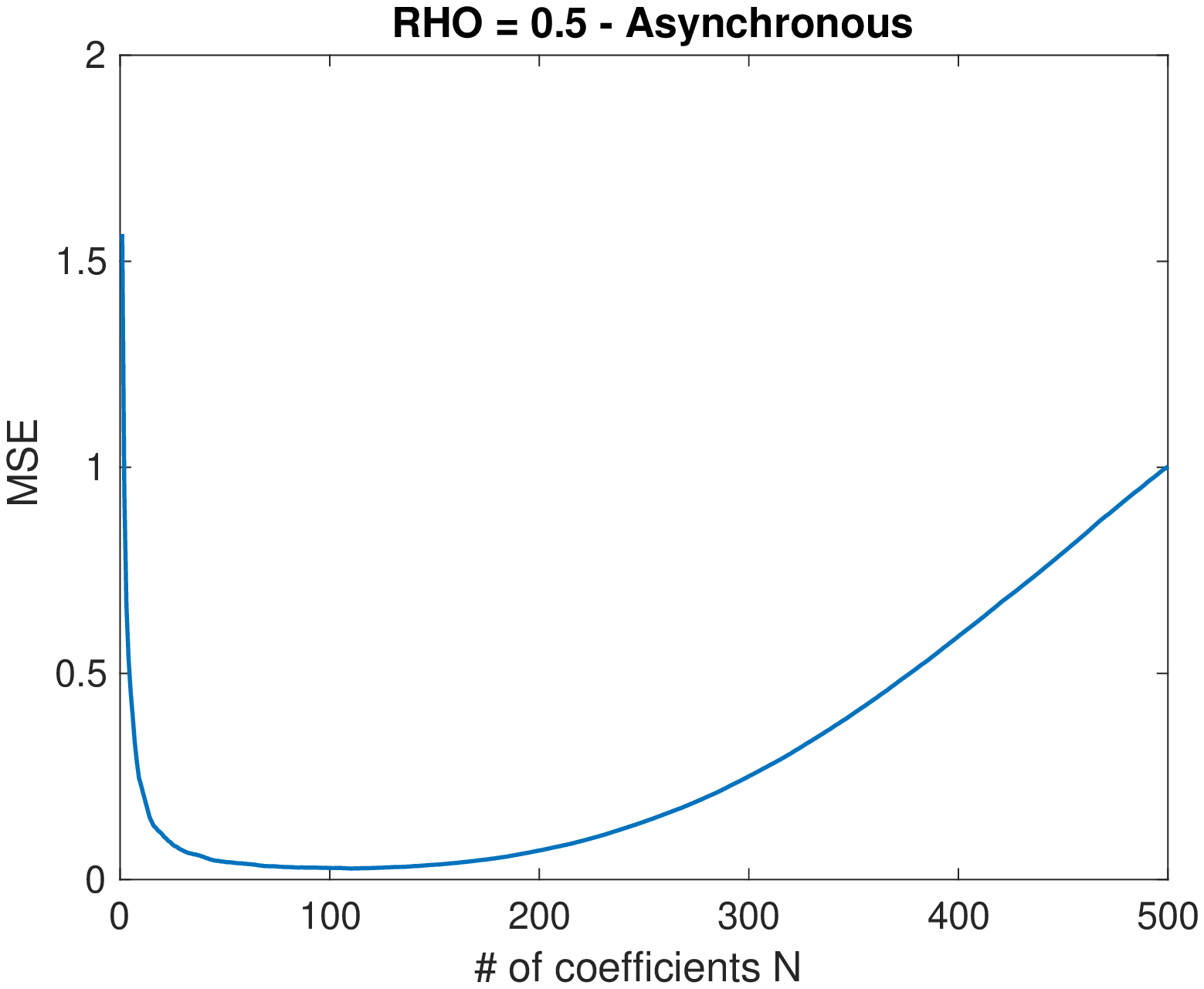}\\
	\includegraphics[width=0.43\linewidth]{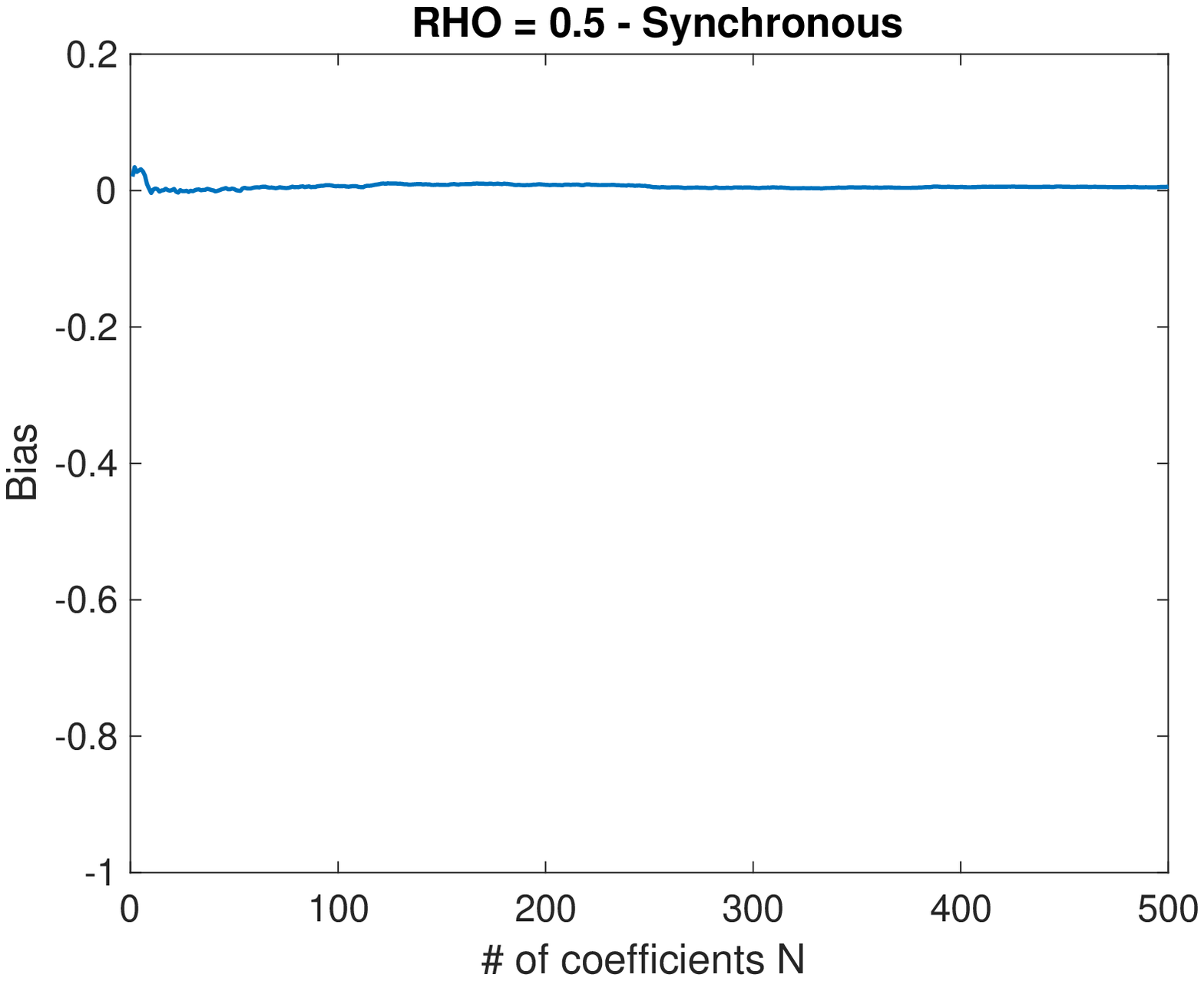}
	\includegraphics[width=0.43\linewidth]{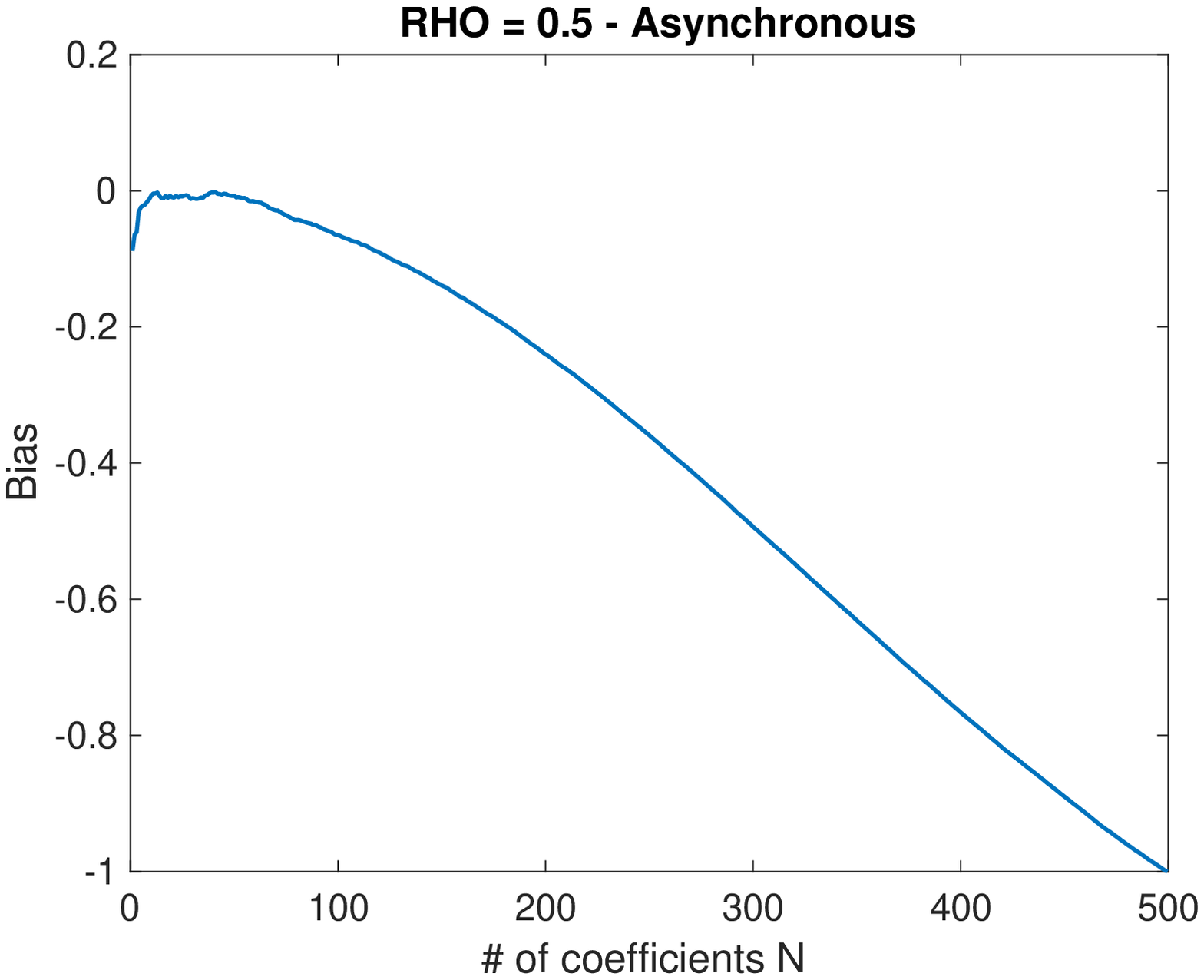}
	\caption{Mean square error (first row) and bias (second row) for $\rho=0.5$.}
	\label{fig:as05}
\end{figure}

\begin{figure}[htpb!]
	\centering
	\includegraphics[width=0.43\linewidth]{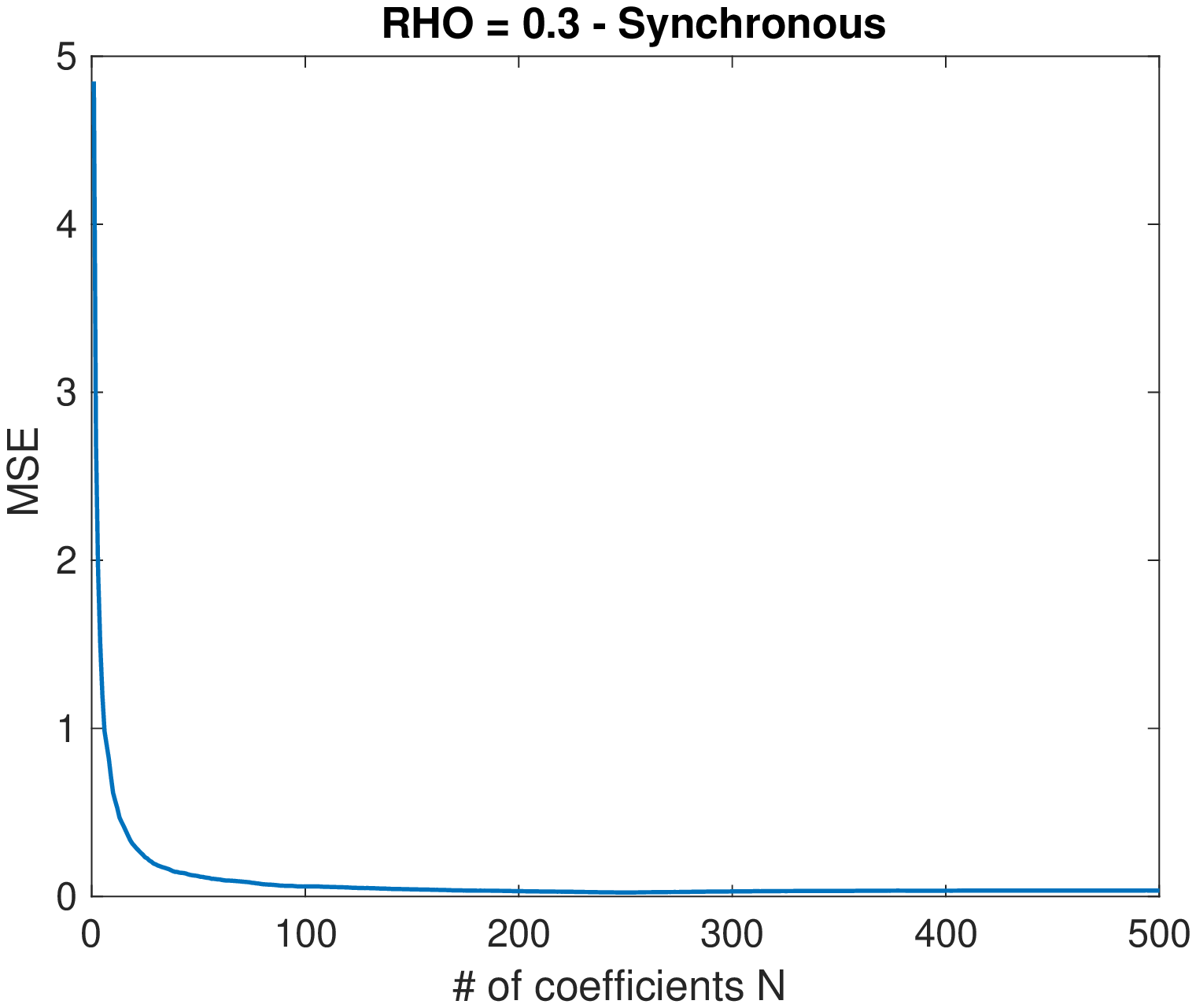}
	\includegraphics[width=0.43\linewidth]{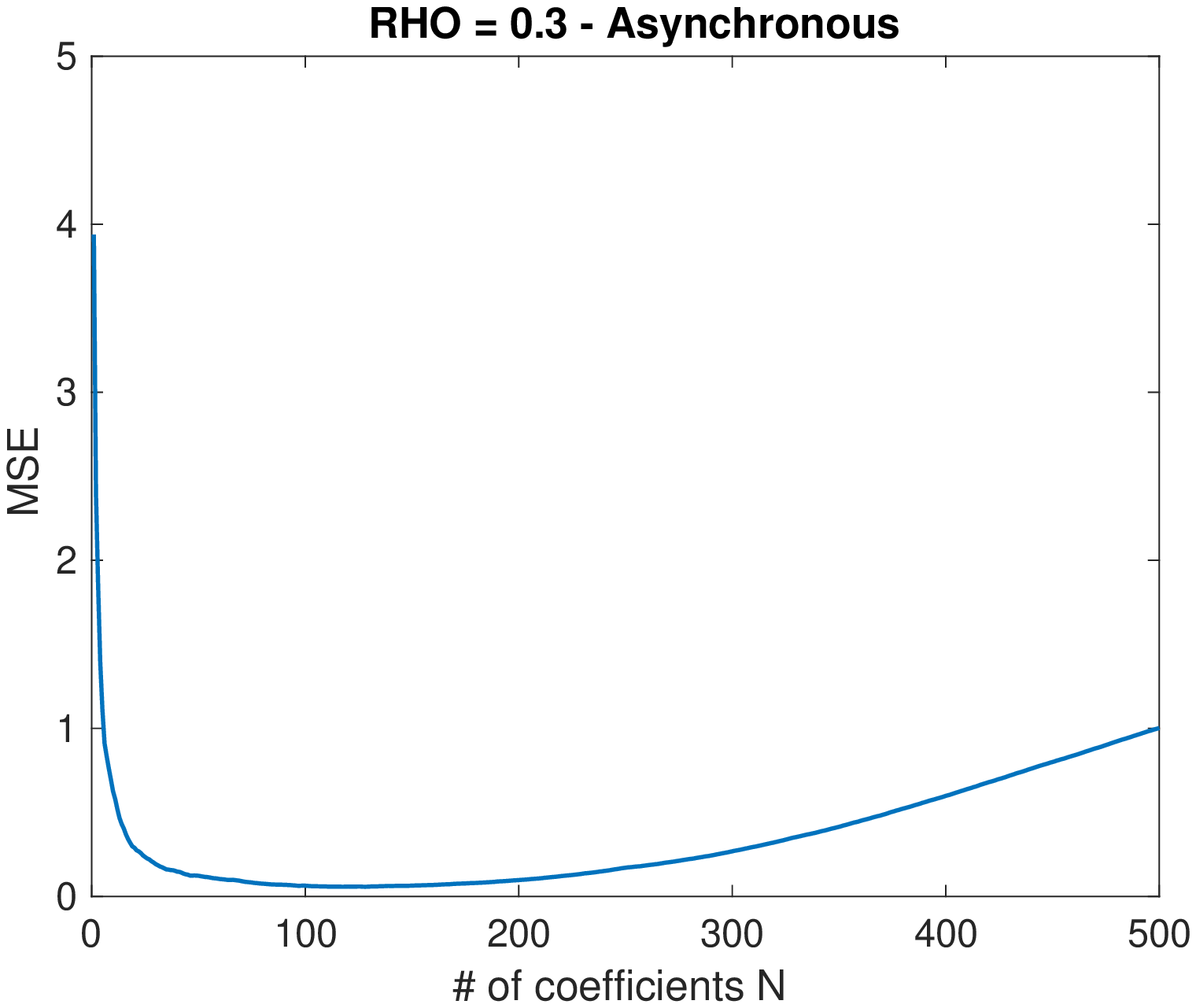}\\
	\includegraphics[width=0.43\linewidth]{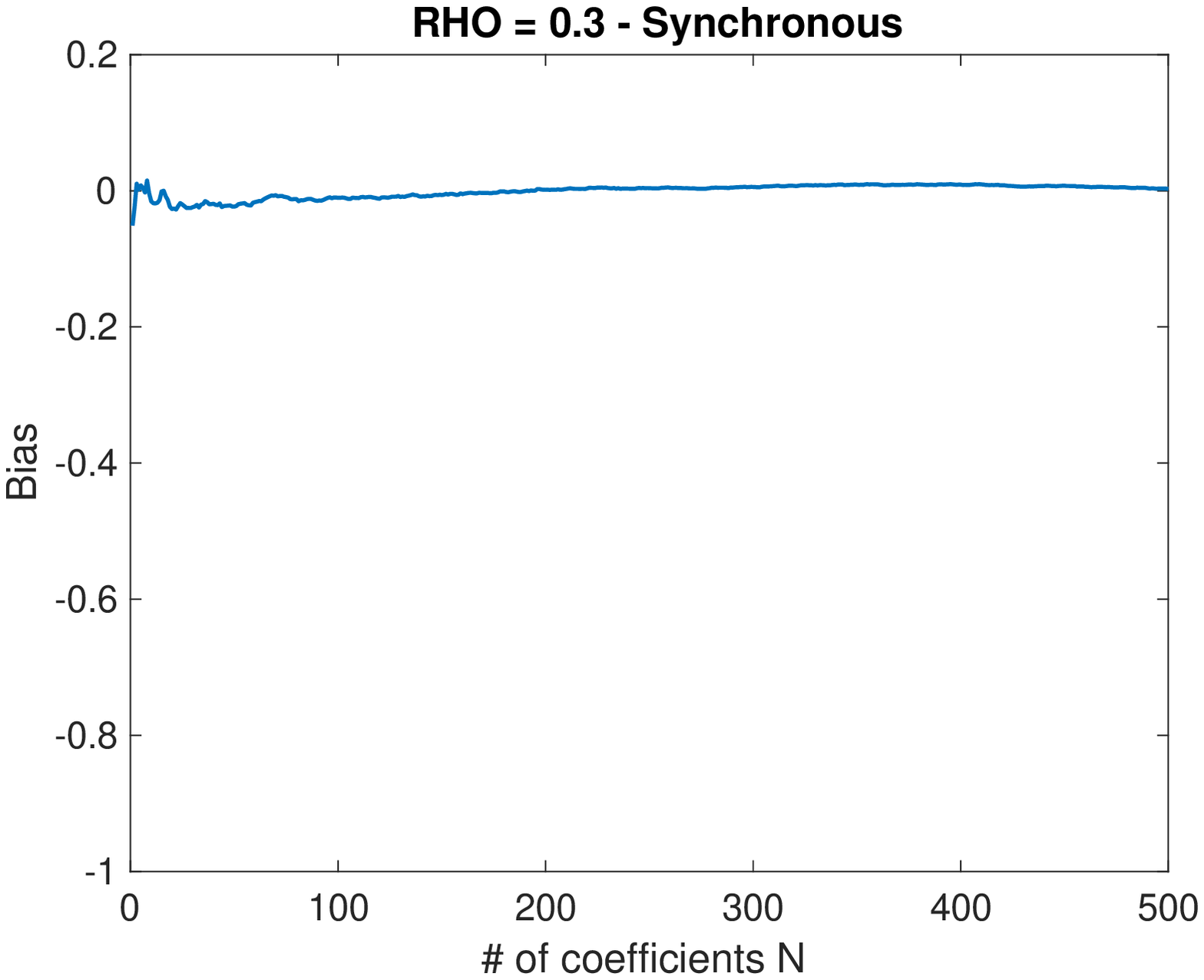}
	\includegraphics[width=0.43\linewidth]{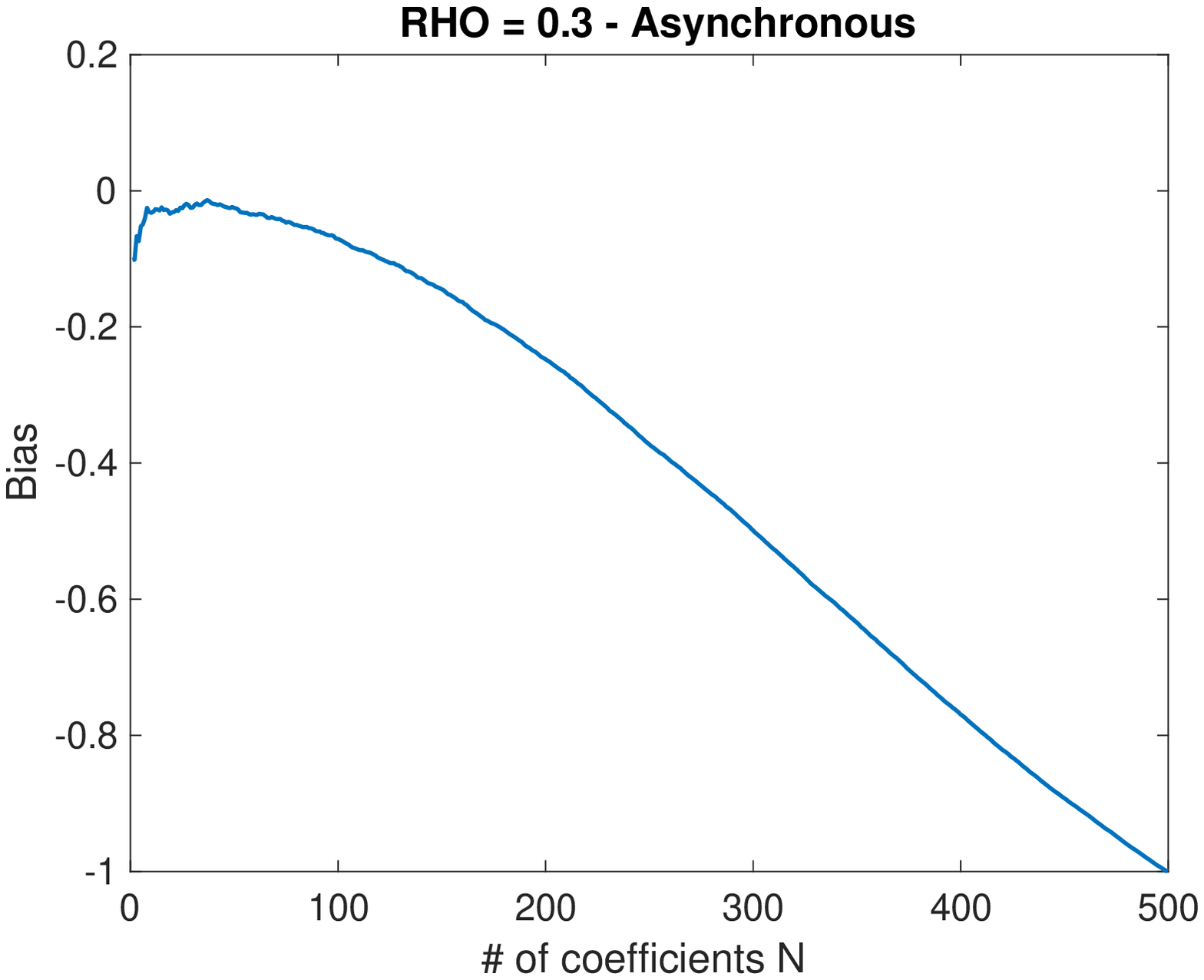}
	\caption{Mean square error (first row) and bias (second row) for $\rho=0.3$.}
	\label{fig:as03}
\end{figure}

\begin{figure}[htpb!]
	\centering
	\includegraphics[width=0.43\linewidth]{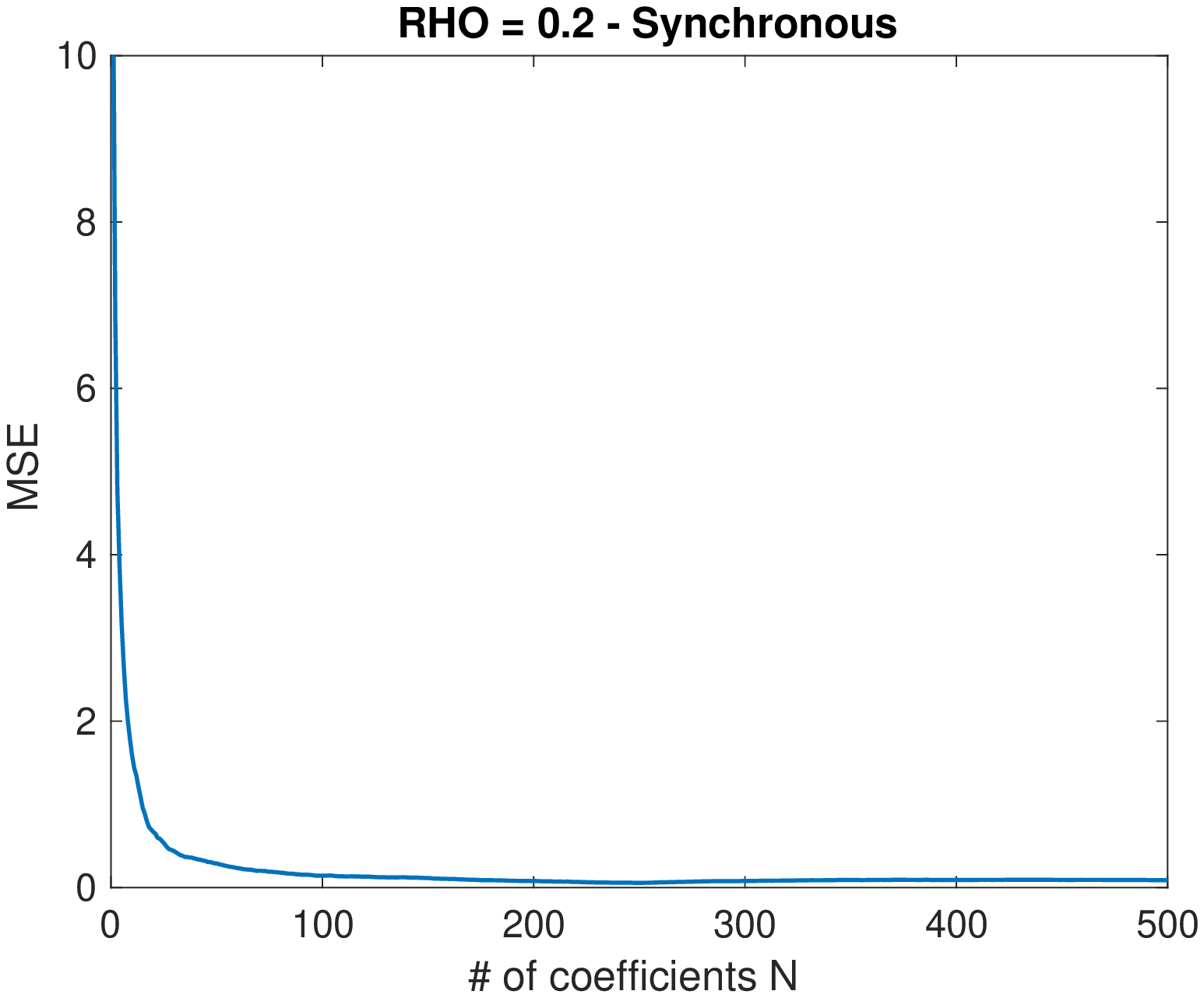}
	\includegraphics[width=0.43\linewidth]{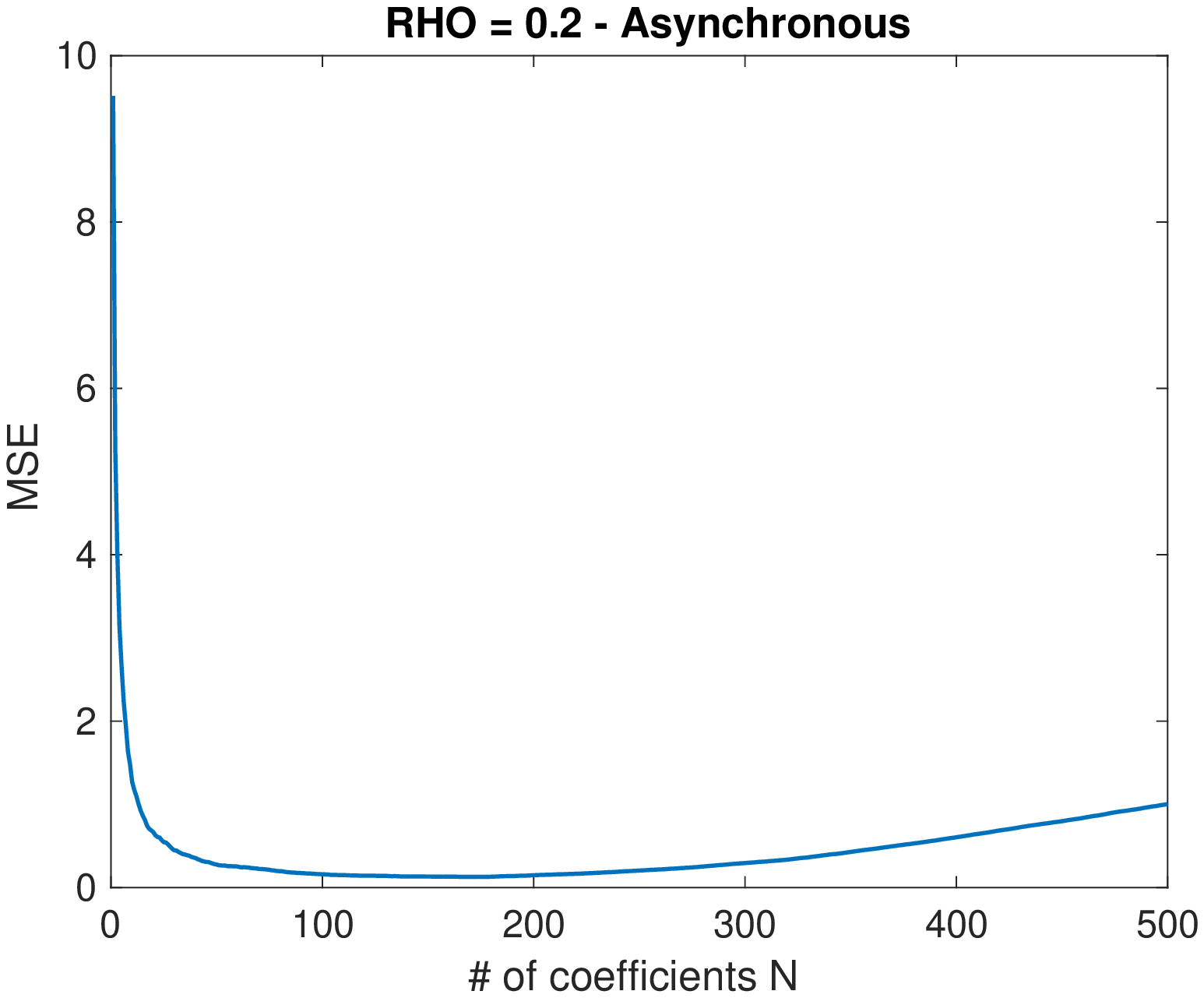}\\
	\includegraphics[width=0.43\linewidth]{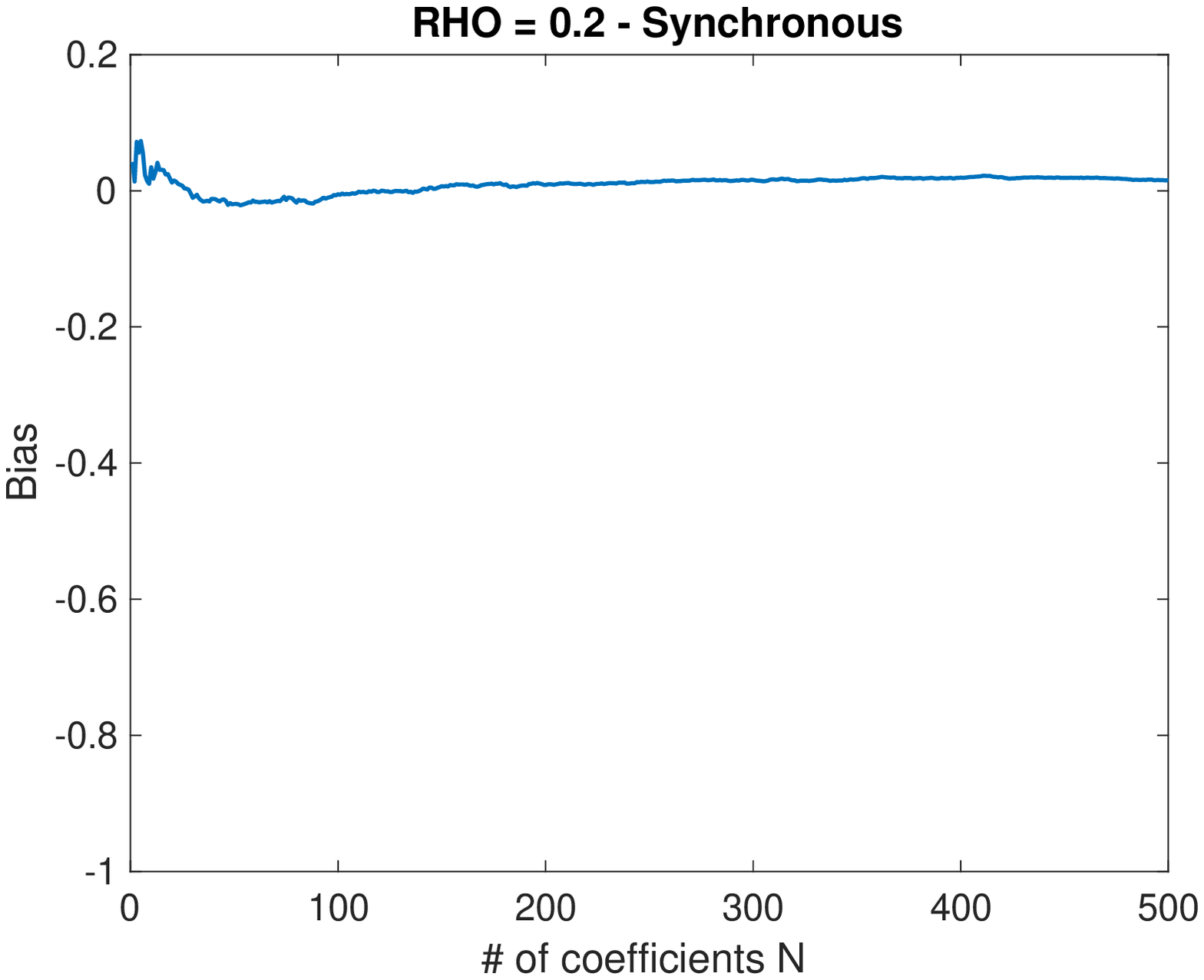}
	\includegraphics[width=0.43\linewidth]{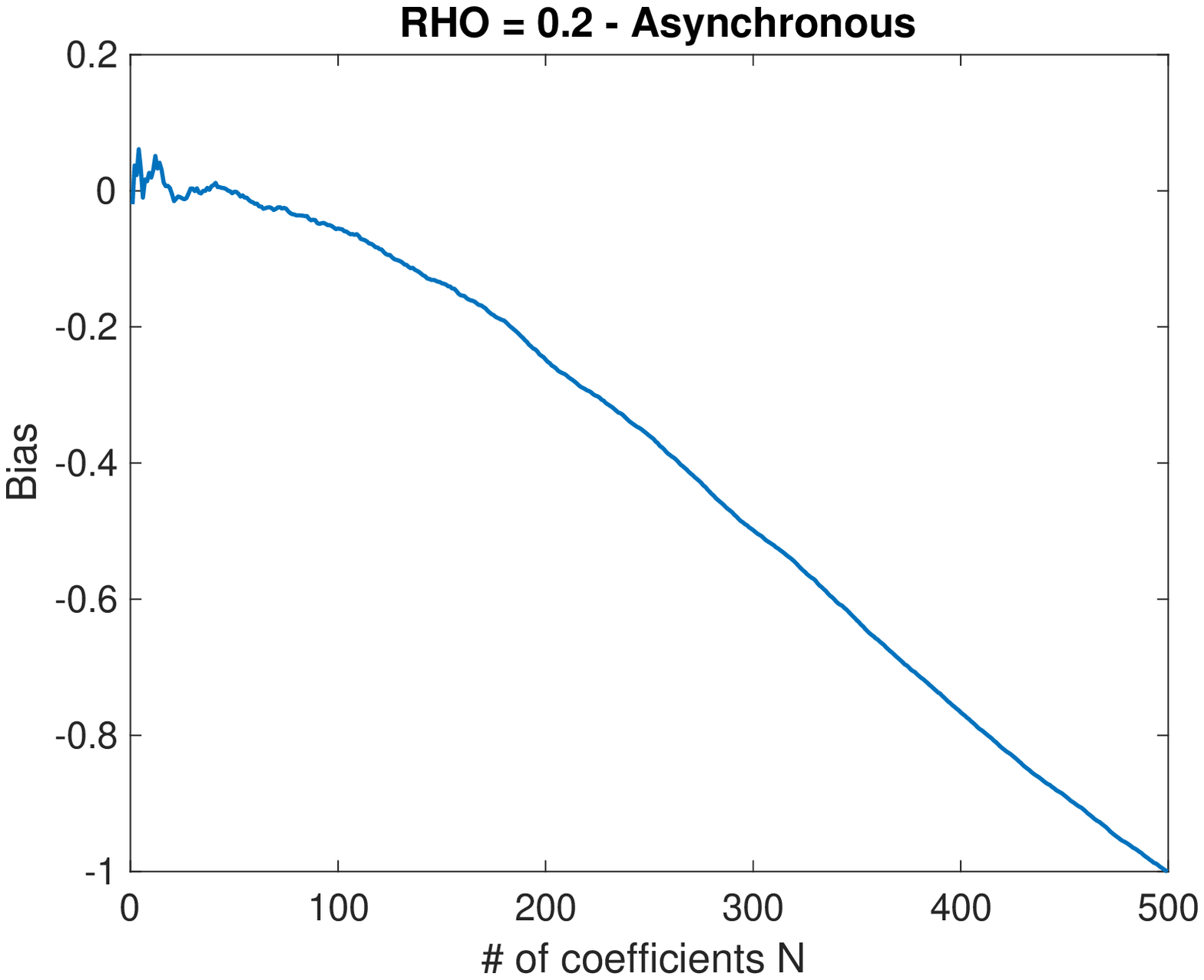}
	\caption{Mean square error (first row) and bias (second row) for $\rho=0.2$.}
	\label{fig:as02}
\end{figure}

Figure \ref{fig:as1} - \ref{fig:as02} show the results of this analysis. In particular, each figure show the plot of the relative mean square error and relative bias of the estimation against the number of Fourier coefficients $N$, for one value of correlation $\rho$, in the two cases of synchronous and asynchronous observations. The results appear to be consistent across the different values of correlation, with the exception of the case in which two perfectly correlated assets are considered.

It is interesting to notice that while in presence of perfect correlation the introduction of asynchronicity seems to have barely any effect, this is not the case when correlation between the two assets decreases. In particular we can notice that setting $\rho \neq 1$ leads to an increased mean square error for large values of $N$ (in particular for values above the Nyquist frequency, as expected, but also in an interval on the left of $n/2$), while the observed increase in th mean square error for the very small values of $N$ seems to be due to a finite sample effect of the change in the correlation. There is, however, a region for which the mean square error appears to be stable when asynchronicity is introduced, and the starting point of this region seems to be close to the value $N=\frac{1}{2}n^{3/4}$, that we found to be optimal in absence of noise in the previous analysis. Moreover, the introduction of asynchronicity seems to induce a significant bias for large values of $N$, and therefore the choice of a smaller $N$ also improves the quality of estimation under this point of view. This exercise shows that choice previously adopted seems to be influenced by the presence of asynchronicity in the data, and is robust to different choices of $\rho$.

While a central limit theorem for the estimator is left for future work, the similar behavior that emerges for different values of correlation appears to be coherent with an asymptotic distribution that does not depend on the estimated quantity itself, as usually happens for high frequency estimators, see, e.g., \citet{ait2014high}.

\subsection{Comparison}\label{sec:com}
	
After having assessed the sensibility of our estimator to the choice of its parameters, in this section we replicate the extensive simulation study adopted in section \ref{sec:choicepar} to evaluate the accuracy and the ability of the proposed PDF estimator in comparison with the ones of two competing estimators that are present in the literature. Focusing on the estimators able to manage asynchronous observations we consider here:

\begin{itemize}
	\item the modified Fourier estimator proposed in this work (PDF);
	\item the smoothed two-scale spot estimator, by \citet{mykland2019algebra} (STS);
	\item the local method of moments spot estimators, by \citet{bibinger2019estimating} (LMM).
\end{itemize}
The kernel-based estimators proposed by \citet{bu2022nonparametric}, while it may be extended to manage irregular and asynchronous observations, relies on specifically tuned shrinkage techniques to impose positive semi-definiteness of the estimation, and is therefore not considered in this analysis.

The analysis is carried out considering a maximum of $d=40$ assets, and the performances of each estimator are evaluated according to the mean integrated square error (MISE) and its relative counterpart (RMISE), defined as
\[
MISE= (Kd^2)^{-1}\sum_{k=1}^K\int_0^{1}\sum_{j,i=1}^d(\hat{V}^{ij}_k(t)-V^{ij}_k(t))^2dt,
\]

\[
RMISE=(Kd^2)^{-1}\sum_{k=1}^K\int_0^{1}\sum_{j,i=1}^d(\hat{V}^{ij}_k(t)-V^{ij}_k(t))^2 /V^{ij}_k(t)^2 dt.
\]

Unreported results show that using a different loss function, in particular the Frobenius norm, the Euclidean norm or the 1-norm of the difference between the estimated and the real spot volatility matrix, does not affect the rankings that emerge from tables \ref{tab:dim}-\ref{tab:het}.
Most importantly, in this analysis we give particular attention to the percentage of symmetric positive semi-definite (psd) variance-covariance matrix that each estimator is able to produce in the different scenarios.

For the LMM estimator, we use the parameters that where found to be optimal in the numerical analysis by \citet{bibinger2019estimating}, while for the STS estimator we choose the parameters in a neighborhood of the one used by \citet{chen2020five}, minimizing the mean square error obtained on auxiliary simulations.

\subsubsection{Absence of noise}\label{sec:sim-nonoise}

In sections \ref{sec:sim-nonoise} and \ref{sec:sim-noise} we report the results of the comparison, in terms of MISE and \% of psd estimations produced by the three competing estimators, when the efficient price process follows the Heston model. In this settings, the results in terms of RMISE are analogous. We begin considering noise-free data, and focusing on two major features of high-frequency covariance matrix estimation: the dimensionality of the matrix and the frequency of observations.
	
It is known that the spectral distribution of a sample covariance matrix depends on the dimension of the matrix itself (or, better, on the ratio between the number of observations and the dimension of the observed random vector); see, e.g, \citet{bai2010spectral} for theory of spectral distribution of random matrices. The impact of matrix dimension is relevant also on applications, for example in principal component analysis (the so-called "curse of dimensionality" and its impact on estimation of eigenvalues, see, e.g., \citet{chen2020five}). However, for what concerns the spot variance estimators considered in this work, nothing is known about the spectral density of the estimated matrices, and the estimated matrices (for the alternative estimators) are not even guaranteed to be positive semi-definite. While addressing the issue of studying in detail the spectral distribution of estimated matrices (that may be useful to build, for example, a finite-sample PCA framework that relies on spot covariance matrices), goes beyond the aim of this work, it is of interest here to assess whether changing the dimension of the problem influence the proportion of positive semi-definite estimations obtained with the considered estimators, hence considering a spectral distribution with possibly negative support. This issue is addressed in all the following exercises considering varying levels of $d$, i.e., an increasing number of assets.

Table \ref{tab:dim} shows the results for data not considered by noise, looking at matrices of dimension $d=2,5,10,15,20,25,30,40$. In this exercise, our modified Fourier estimator is the top performer in terms of mean square error, for any dimension of the volatility matrix, when the efficient price process is observed. The effectiveness of the smoothed two-scale estimator to produce positive semi-definite estimations seems to decrease while the number of assets increase, in particular with $d>20$, while the other two estimators both produce 100\% of psd matrices. For the important role that plays in estimating variance-covariance matrices, and for the influence that it has on the positivity of the estimation, dimensionality will always be taken into consideration in the remaining analysis. For sake of simplicity, the dimension considered in the following are limited to $d=5,10,15,20$.

\begin{table}[htbp!]
	\begin{center}
		\begin{tabular}{c|c|c||c|c}
			\hline\hline
			Estimator & MISE &\% PSD & MISE &\% PSD\\
			\hline\hline
			& \multicolumn{2}{c||}{d=2} & \multicolumn{2}{c}{d=20}\\
			\hline
			PDF &  1.8453e-04 &  100\% & 6.5495e-04 &  100\%\\
			LMM &  5.2131e-04 & 100\% &  1.6008e-03 &  100\% \\
			STS &  8.2259e-04 &  100\% & 2.5759e-03 &  100\%\\
			\hline
			& \multicolumn{2}{c||}{d=5} & \multicolumn{2}{c}{d=25}\\
			\hline
			PDF & 5.7567e-04 &  100\% & 6.1793e-04  &  100\%\\
			LMM & 1.5514e-03 & 100\% & 1.5090e-03 &  100\% \\
			STS & 2.3805e-03 &  100\% & 2.4401e-03 &  99.20\%\\
			\hline
			& \multicolumn{2}{c||}{d=10} & \multicolumn{2}{c}{d=30}\\
			\hline
			PDF & 8.3119e-04 &  100\% & 6.8819e-04 &  100\%\\
			LMM & 2.0574e-03 & 100\% & 1.6779e-03 &  100\% \\
			STS & 3.3500e-03 &  100\% & 2.7116e-03 &  98.56\%\\
			\hline
			& \multicolumn{2}{c||}{d=15} & \multicolumn{2}{c}{d=40}\\
			\hline
			PDF & 6.9991e-04 &  100\% & 6.3139e-04  &  100\%\\
			LMM & 1.7120e-03 & 100\% & 1.5390e-03 &  100\% \\
			STS & 2.7634e-03 &  100\% & 2.4887e-03  &  33.10\%\\
			\hline
		\end{tabular}
	\caption{Accuracy (MISE) and \% of psd matrix produced by each estimator, when the dimension $d$ of $V$ increase.}\label{tab:dim}
	\end{center}
\end{table}

In a setting of absence of noise we take into consideration the presence of different sampling schemes for the data. To do so, we study the changes in the performances when the average time between two consecutive observations increases; in particular we extract the observation from the simulated trajectories according to Poisson processes that produce on average one observation every 15, 30 and 40 seconds. Table \ref{tab:as} shows that, still maintaining an edge in terms of MISE with respect to the competitors, the PDF estimator is the only one that is able to produce psd estimations in 100\% of the cases, while both the STS and the LMM estimator may fails with increased frequency when $\bar{\Delta}t$ increase, even though the impact of this kind of changes seems to be quite small in terms of percentage of positive estimation obtained. It also seems that the accuracy of all the estimators decreases with higher values of $\bar{\Delta}t$; this is of course in line with the fact that the consistency of these estimators is an asymptotic property.

\begin{table}[htbp!]
	\begin{center}
		\begin{tabular}{c|c|c||c|c||c|c}
			\hline\hline
			Estimator & MISE &\% PSD & MISE &\% PSD & MISE &\% PSD\\
			\hline\hline
			& \multicolumn{2}{c||}{d=5, $\bar{\Delta} t=15$ } & \multicolumn{2}{c||}{d=5, $\bar{\Delta} t=30$} & \multicolumn{2}{c}{d=5, $\bar{\Delta} t=40$}\\
			\hline
			PDF & 7.7661e-04 &  100\% & 1.2284e-03 &  100\%& 1.5871e-03 &  100\%\\
			LMM & 1.5937e-03 &  100\% & 2.0398e-03 &  100\%& 2.3025e-03  &  99.83\%\\
			STS & 2.5185e-03 &  100\% & 3.4587e-03 &  100\%& 4.3867e-03  &  100\%\\
			\hline
			& \multicolumn{2}{c||}{d=10, $\bar{\Delta} t=15$} & \multicolumn{2}{c||}{d=10, $\bar{\Delta} t=30$}& \multicolumn{2}{c}{d=10, $\bar{\Delta} t=40$}\\
			\hline
			PDF & 1.1008e-03 &  100\% & 1.7661e-03 &  100\%& 2.2329e-03 &  100\%\\
			LMM & 2.2549e-03 & 100\% & 2.7865e-03 &  100\% & 3.0841e-03 &  99.66\%\\
			STS & 3.4760e-03 &  100\% & 4.8142e-03 &  100\% & 6.4107e-03  &  100\%\\
			\hline
			& \multicolumn{2}{c||}{d=15, $\bar{\Delta} t=15$} & \multicolumn{2}{c||}{d=15, $\bar{\Delta} t=30$}& \multicolumn{2}{c}{d=15, $\bar{\Delta} t=40$}\\
			\hline
			PDF & 9.0624e-04 &  100\% & 1.4832e-03  &  100\% & 1.8593e-03  &  100\%\\
			LMM & 1.8481e-03 & 100\% & 2.2859e-03 &  100\% & 2.5646e-03 &  98.62\%\\
			STS & 2.8926e-03 &  100\% & 4.0185e-03  &  99.98\% & 5.3583e-03  & 99.75 \%\\
			\hline
			& \multicolumn{2}{c||}{d=20, $\bar{\Delta} t=15$} & \multicolumn{2}{c||}{d=20, $\bar{\Delta} t=30$}& \multicolumn{2}{c}{d=20, $\bar{\Delta} t=40$}\\
			\hline
			PDF & 8.5236e-04 &  100\% & 1.3852e-03  &  100\% & 1.8593e-03  &  100\%\\
			LMM & 1.7170e-03 & 100\% & 2.1356e-03 &  99.96\% & 2.5646e-03 &  98.62\%\\
			STS & 2.6941e-03 &  99.98\% & 3.7669e-03  &  98.83\% & 5.3583e-03  & 99.08 \%\\
		\end{tabular}
	\end{center}
\caption{Accuracy and \% of psd matrix produced by each estimator, when the average time between consecutive observations $\bar{\Delta}t$ changes.}\label{tab:as}
\end{table}

\subsubsection{Data contaminated by MMN}\label{sec:sim-noise}

In this section we run our comparison considering the noise specification described in section \ref{sec:nm}. It is useful to remark that, while the LMM estimator entails an explicit noise correction and the STS estimator relies on pre-averaging of the data, for the proposed PDF estimator, in line with the original Fourier estimator of spot volatility, there is no need to manipulate data or correct the estimator to manage the presence of noise, but it is sufficient to cut the frequency $N$, as shown in section \ref{sec:choicepar}.

Table \ref{tab:rou} shows that the presence of rounding seems not to affect significantly the accuracy of the estimators and the positive semi-definiteness of the estimations. This effect may be due to the scheme adopted to simulate irregularly sampled data, that implies a subsampling with respect to the rounded simulated series, reducing the intensity of this source of noise.

\begin{table}[htbp!]
	\begin{center}
		\begin{tabular}{c|c|c||c|c}
			\hline\hline
			Estimator & MISE &\% PSD & MISE &\% PSD\\
			\hline\hline
			& \multicolumn{2}{c||}{d=5, r=0.01} & \multicolumn{2}{c}{d=5, r=0.05}\\
			\hline
			PDF & 5.6886e-04 &  100\% & 5.8434e-04  &  100\%\\
			LMM & 1.4588e-03 & 100\% & 1.4744e-03 &  100\% \\
			STS & 2.4015e-03 &  100\% & 2.4342e-03 &  100\%\\
			\hline
			& \multicolumn{2}{c||}{d=10, r=0.01} & \multicolumn{2}{c}{d=10, r=0.05}\\
			\hline
			PDF & 8.2010e-04 &  100\% & 8.2171e-04 &  100\%\\
			LMM & 2.0337e-03 & 100\% & 2.0445e-03 &  100\% \\
			STS & 3.3250e-03 &  100\% & 3.3612e-03 &  100\%\\
			\hline
			& \multicolumn{2}{c||}{d=15, r=0.01} & \multicolumn{2}{c}{d=15, r=0.05}\\
			\hline
			PDF & 6.7300e-04 &  100\% & 6.7597e-04  &  100\%\\
			LMM & 1.6870e-03 & 100\% & 1.6806e-03 &  100\% \\
			STS & 2.7879e-03 &  100\% & 2.7996e-03  &  100\%\\
			\hline
			& \multicolumn{2}{c||}{d=20, r=0.01} & \multicolumn{2}{c}{d=20, r=0.05}\\
			\hline
			PDF & 6.3197e-04 &  100\% & 6.3562e-04  &  100\%\\
			LMM & 1.5806e-03 & 100\% & 1.5791e-03 &  100\% \\
			STS & 2.6104e-03 &  100\% & 2.6063e-03  &  100\%\\
			\hline
		\end{tabular}
	\end{center}
\caption{Accuracy and \% of psd matrix produced by each estimator, when a rounding of 1 or 5 cents is present.}\label{tab:rou}
\end{table}

Table \ref{tab:n-iid} shows that i.i.d. noise, especially with high $\sigma^2_\eta$, is able to negatively affect the ability of the LMM and STS estimators to produce psd estimations, with a stronger impact as the dimension of the estimated matrix grows. Also, the accuracy of all the estimators deteriorates with higher noise, with the PDF confirmed as the top performer also in this scenario. Since i.i.d. noise, form a market microstructure perspective, is usually linked to the presence bid-ask spread as modelled, e.g., as in \citet{roll1984simple}, and being the bid-ask spread usually related to the liquidity of an asset, the ability of managing this kind of noise may be regarded as the ability to estimate correctly covariance also for illiquid assets.

\begin{table}[htbp!]
	\hspace{-1cm}
		\begin{tabular}{c|c|c||c|c||c|c||c|c}
			\hline\hline
			Estimator & MISE &\% PSD & MISE &\% PSD & MISE &\% PSD & MISE &\% PSD\\
			\hline\hline
			& \multicolumn{2}{c||}{d=5, $\sigma_{\eta}=1$ } & \multicolumn{2}{c||}{d=5, $\sigma_{\eta}=1.5$} & \multicolumn{2}{c||}{d=5, $\sigma_{\eta}=2$} & \multicolumn{2}{c}{d=5, $\sigma_{\eta}=2.5$}\\
			\hline
			PDF & 5.8957e-04 &  100\% & 1.2411e-03 &  100\%& 2.1739e-03 &  100\%& 3.5475e-03 & 100\%\\
			LMM & 1.4774e-03 &  100\% & 2.0615e-03 &  100\%& 3.2557e-03  &  99.91\% & 5.5029e-03 & 98.18\%\\
			STS & 2.4258e-03 &  100\% & 3.0744e-03 &  100\%& 3.9064e-03  &  99.95\%& 5.1760e-03 & 97.85\%\\
			\hline
			& \multicolumn{2}{c||}{d=10, $\sigma_{\eta}=1$} & \multicolumn{2}{c||}{d=10, $\sigma_{\eta}=1.5$}& \multicolumn{2}{c||}{d=10, $\sigma_{\eta}=2$} & \multicolumn{2}{c}{d=10, $\sigma_{\eta}=2.5$}\\
			\hline
			PDF & 8.3152e-04 &  100\% & 1.7158e-03 &  100\%& 2.4388e-03 &  100\%& 3.5411e-03& 100\%\\
			LMM & 2.0447e-03 & 100\% & 2.6918e-03 &  100\% & 3.6557e-03 &  99.56\%& 5.5120e-03& 91.14\%\\
			STS & 3.3806e-03 &  100\% & 4.3057e-03 &  99.83\% & 5.4130e-03  &  90.31\%& 7.0991e-03 &52.19 \%\\
			\hline
			& \multicolumn{2}{c||}{d=15, $\sigma_{\eta}=1$} & \multicolumn{2}{c||}{d=15, $\sigma_{\eta}=1.5$}& \multicolumn{2}{c||}{d=15, $\sigma_{\eta}=2$} & \multicolumn{2}{c}{d=15, $\sigma_{\eta}=2.5$}\\
			\hline
			PDF & 6.8032e-04 &  100\% & 1.4012e-03  &  100\% & 1.8502e-03  &  100\%& 2.5615e-03 & 100\%\\
			LMM & 1.6911e-03 & 100\% & 2.1709e-03 &  99.98\% & 2.8183e-03 &  99.20\%& 3.9896e-03 & 80.66\%\\
			STS & 2.8036e-03 & 99.87\% & 3.5990e-03 & 91.42\% & 5.8970e-03  & 34.15 \%& & 26.37\%\\
			\hline
			& \multicolumn{2}{c||}{d=20, $\sigma_{\eta}=1$} & \multicolumn{2}{c||}{d=20, $\sigma_{\eta}=1.5$}& \multicolumn{2}{c||}{d=20, $\sigma_{\eta}=2$} & \multicolumn{2}{c}{d=20, $\sigma_{\eta}=2.5$}\\
			\hline
			PDF & 6.3538e-04 &  100\% & 1.3026e-03  &  100\% & 1.6698e-03  &  100\%& 2.2492e-03& 100\%\\
			LMM & 1.5694e-03 & 100\% & 2.0025e-03 &  99.96\% & 2.5535e-03 &  97.44\%& 3.5224e-03& 64.45\%\\
			STS & 2.6111e-03 & 97.51\% & 3.3587e-03 & 42.48\% & 4.1644e-03  & 11.71 \%&5.4918e-03 & 0.50\%\\
			\hline
		\end{tabular}
\caption{Accuracy and \% of psd matrix produced by each estimator, when the data is contaminated by i.i.d. noise.}\label{tab:n-iid}
\end{table}

Table \ref{tab:aut} shows that autocorrelation is able to significantly effect the ability of the STS estimators to produce psd matrices, in particular with low values of $\theta_\eta$; also the LMM estimator is affected, even if more slightly, Low values of $\theta_\eta$ also reduce the accuracy of all the three estimators, while maintaining their ranking unchanged.

\begin{table}[htbp!]
	\begin{center}
		\begin{tabular}{c|c|c||c|c||c|c}
			\hline\hline
			Estimator & MISE &\% PSD & MISE &\% PSD & MISE &\% PSD\\
			\hline\hline
			& \multicolumn{2}{c||}{d=5, $\theta_{\eta}=0.2$ } & \multicolumn{2}{c||}{d=5, $\theta_{\eta}=0.3$} & \multicolumn{2}{c}{d=5, $\theta_{\eta}=0.4$}\\
			\hline
			PDF & 5.6169e-03 &  100\% & 2.8806e-03 &  100\%& 2.1414e-03 &  100\%\\
			LMM & 1.0856e-02 & 	100\% & 4.8943e-03 &  100\%& 3.3544e-03  & 100\%\\
			STS & 8.5059e-03 &  97.31\% & 5.1553e-03 &  99.08\%& 4.1302e-03  &  99.92\%\\
			\hline
			& \multicolumn{2}{c||}{d=10, $\theta_{\eta}=0.2$} & \multicolumn{2}{c||}{d=10, $\theta_{\eta}=0.3$}& \multicolumn{2}{c}{d=10, $\theta_{\eta}=0.4$}\\
			\hline
			PDF & 5.1481e-03 &  100\% & 3.0019e-03 &  100\%& 2.3519e-03 &  100\%\\
			LMM & 9.7818e-03 & 100\% & 5.0765e-03 &  99.77\% & 3.7482e-03 &  100\%\\
			STS & 1.0814e-02 &  60.36\% & 6.8148e-03 &  90.08\% & 5.6242e-03  &  91.37\%\\
			\hline
			& \multicolumn{2}{c||}{d=15, $\theta_{\eta}=0.2$} & \multicolumn{2}{c||}{d=15, $\theta_{\eta}=0.3$}& \multicolumn{2}{c}{d=15, $\theta_{\eta}=0.4$}\\
			\hline
			PDF & 3.5769e-03 &  100\% & 2.2202e-03  &  100\% & 1.8000e-03  &  100\%\\
			LMM & 6.6905e-03 & 92.90\% & 3.7483e-03 &  95.98\% & 2.8798e-03 &  96.80\%\\
			STS & 8.7234e-03 & 3.70\% & 5.6586e-03 &  29.92\% & 4.6585e-03  & 35.03\%\\
			\hline
			& \multicolumn{2}{c||}{d=20, $\theta_{\eta}=0.2$} & \multicolumn{2}{c||}{d=20, $\theta_{\eta}=0.3$}& \multicolumn{2}{c}{d=20, $\theta_{\eta}=0.4$}\\
			\hline
			PDF & 3.0654e-03 &  100\% & 1.9632e-03  &  100\% & 1.6249e-03  &  100\%\\
			LMM & 5.6948e-03 & 93.90\% & 3.3182e-03 &  92.98\% & 2.6127e-03 &  95.48\%\\
			STS & 8.0435e-03 & 4.87\% & 5.2368e-03 & 9.65\% & 4.3666e-03  & 8.23\%\\
			\hline
		\end{tabular}
	\end{center}
\caption{Accuracy and \% of psd matrix produced by each estimator, when the data is contaminated by autocorrelated noise.}\label{tab:aut}
\end{table}		

Table \ref{tab:corr} shows that introducing correlation between $E$ and $W$, i.e. between the Brownian motions that drives the noise and the one that drives the dynamic of the efficient price, does not seem to influence particularly the positive semi-definiteness of the estimations. Moreover, it seems that the accuracy on slightly increase for all the estimators, with respect to the baseline i.i.d. case; this is due to the fact that a negative autocorrelation between efficient price and noise may induce a bias with opposite sign whit respect to the bias induced by uncorrelated noise; this is usually observed using volatility signature plots, as done in \citet{hansen2006realized}.

\begin{table}[htbp!]
	\begin{center}
		\begin{tabular}{c|c|c||c|c||c|c}
			\hline\hline
			Estimator & MISE &\% PSD & MISE &\% PSD & MISE &\% PSD\\
			\hline\hline
			& \multicolumn{2}{c||}{d=5, $\rho_{\eta}=-0.1$ } & \multicolumn{2}{c||}{d=5, $\rho_{\eta}=-0.3$} & \multicolumn{2}{c}{d=5, $\rho_{\eta}=-0.5$}\\
			\hline
			PDF & 2.0352e-03 &  100\% & 1.9671e-03 &  100\%& 1.9375e-03
			 &  100\%\\
			LMM & 3.1654e-03 & 	100\% & 3.0462e-03 &  100\%& 2.9484e-03  & 100\%\\
			STS & 3.9369e-03 &  99.81\% & 3.8525e-03 &  99.96\%& 3.8145e-03  &  99.92\%\\
			\hline
			& \multicolumn{2}{c||}{d=10, $\rho_{\eta}=-0.1$} & \multicolumn{2}{c||}{d=10, $\rho_{\eta}=-0.3$}& \multicolumn{2}{c}{d=10, $\rho_{\eta}=-0.5$}\\
			\hline
			PDF & 2.2896e-03 &  100\% & 2.3440e-03 &  100\%& 2.2369e-03 &  100\%\\
			LMM & 3.5607e-03 & 99.94\% & 3.4779e-03 &  99.92\% & 3.4024e-03 &  99.98\%\\
			STS & 5.3662e-03 &  91.56\% & 5.2421e-03 &  94.05\% & 5.1648e-03  &  95.49\%\\
			\hline
			& \multicolumn{2}{c||}{d=15, $\rho_{\eta}=-0.1$} & \multicolumn{2}{c||}{d=15, $\rho_{\eta}=-0.3$}& \multicolumn{2}{c}{d=15, $\rho_{\eta}=-0.5$}\\
			\hline
			PDF & 1.7608e-03 &  100\% & 1.7866e-03  &  100\% & 1.7255e-03  &  100\%\\
			LMM & 2.7502e-03 & 98.62\% & 2.7090e-03 &  99.44\% & 2.6483e-03 &  99.79\%\\
			STS & 4.4396e-03 & 37.17\% & 4.3377e-03 & 44.13\% & 4.2908e-03  & 53.57\%\\
			\hline
			& \multicolumn{2}{c||}{d=20, $\rho_{\eta}=-0.1$} & \multicolumn{2}{c||}{d=20, $\rho_{\eta}=-0.3$}& \multicolumn{2}{c}{d=20, $\rho_{\eta}=-0.5$}\\
			\hline
			PDF & 1.6065e-03 &  100\% & 1.6253e-03  &  100\% & 1.5582e-03  &  100\%\\
			LMM & 2.5180e-03 & 98.35\% & 2.4725e-03 &  98.89\% & 2.3988e-03 &  95.84\%\\
			STS & 4.1353e-03 & 0.82\% & 4.0389e-03 & 2.92\% & 3.9787e-03  & 5.43\%\\
			\hline
		\end{tabular}
	\end{center}
\caption{Accuracy and \% of psd matrix produced by each estimator, when the data is contaminated by noise correlated with the efficient price process.}\label{tab:corr}
\end{table}

Table \ref{tab:het} shows the results for the last specification of noise that we consider, with noise autocovariance, correlation with the efficient price and time-varying noise variance. Our results show that, in this setting, both the LMM and the STS estimators may have difficulties in reaching satisfactory percentages of psd estimations, depending on the intensity of the microstructure component. We confirm once again the ability of the PDF estimator to produce variance-covariance matrix with the desired property, and with relatively low estimation error.

\begin{table}[htbp!]
	\begin{center}
		\begin{tabular}{c|c|c||c|c||c|c}
			\hline\hline
			Estimator & MISE &\% PSD & MISE &\% PSD & MISE &\% PSD\\
			\hline\hline
			& \multicolumn{2}{c||}{d=5,  $\bar{\sigma}_\eta=3$} & \multicolumn{2}{c||}{d=5, $\bar{\sigma}_\eta=3.5$} & \multicolumn{2}{c}{d=5, $\bar{\sigma}_\eta=4$}\\
			\hline
			PDF & 4.7998e-03 & 100\% & 7.8667e-03  &  100\% & 1.3212e-02 & 100\%\\
			LMM & 6.8314e-03 & 100\% & 1.1376e-02 &  98.81\% & 1.8293e-02 & 93.86\% \\
			STS & 8.6164e-03 & 84.19\% & 1.3869e-02 &  71.84\%& 2.0190e-02 &61.96\%\\
			\hline
			& \multicolumn{2}{c||}{d=10, $\bar{\sigma}_\eta=3$} & \multicolumn{2}{c||}{d=10, $\bar{\sigma}_\eta=3.5$}& \multicolumn{2}{c}{d=10, $\bar{\sigma}_\eta=4$}\\
			\hline
			PDF & 4.5443e-03 &  100\% & 6.9249e-03 &  100\%&1.1111e-02 &100\%\\
			LMM & 6.7069e-03 & 99.89\% & 1.0364e-02 &  95.89\% &1.6497e-02 &83.48\% \\
			STS & 1.1635e-02 &  52.86\% & 1.7614e-02 &  36.65\%&2.6044e-02 & 30.51\%\\
			\hline
			& \multicolumn{2}{c||}{d=15, $\bar{\sigma}_\eta=3$} & \multicolumn{2}{c||}{d=15, $\bar{\sigma}_\eta=3.5$}& \multicolumn{2}{c}{d=15, $\bar{\sigma}_\eta=4$}\\
			\hline
			PDF & 3.2065e-03 &  100\% & 4.7542e-03  &  100\% &7.2763e-03 &100\% \\
			LMM & 4.8670e-03 & 99.37\% & 7.2946e-03 &  89.26\%  &1.1172e-02 & 68.90\%\\
			STS & 9.4853e-03 &  35.25\% & 1.4449e-02  &  24.56\% &2.1268e-02 &19.98\%\\
			\hline
			& \multicolumn{2}{c||}{d=20, $\bar{\sigma}_\eta=3$} & \multicolumn{2}{c||}{d=20, $\bar{\sigma}_\eta=3.5$}& \multicolumn{2}{c}{d=20, $\bar{\sigma}_\eta=4$}\\
			\hline
			PDF & 2.7842e-03 &  100\% & 4.0233e-03  &  100\% & 6.0704e-03 &100\%\\
			LMM & 4.2737e-03 & 98.14\% & 6.2574e-03 &  80.82\% &9.4823e-03 & 54.03\%\\
			STS & 8.7501e-03 &  25.69\% & 1.3279e-02  &  16.33\% &1.9663e-02
			 &11.38\%\\
			\hline
		\end{tabular}
	\end{center}
\caption{Accuracy and \% of psd matrix produced by each estimator, when the data is contaminated by heteroskedastic noise process.}\label{tab:het}
\end{table}

\subsection{Alternative volatility models}\label{sec:sim-almod}

In the previous sections we have exposed the results of our comparison for what concerns simulation of the efficient price process made with an Heston model. Even though the error produced by the three estimators may be different changing the simulation model behind our analysis, and in particular the differences in MISE between the PDF and the LMM estimators are reduced when using the SVF2 or the Rough Heston model, the results are substantially confirmed: the PDF estimator remain the only one able to consistently produce positive semi-definite estimations, and is the best performer in terms of mean square error in almost any scenario. Table \ref{tab:models} shows the percentage of psd estimations obtained under the alternative volatility models, in absence of microstructure noise. We can see that, in this exercise, it seems that, moving to the SV1F, to the SV2F or Rough Heston, does not influence significantly the ability of the estiamtors of producing positive matrices, and the LMM estimator shows the same ability of producing psd matrices than the PDF estimator. More extensive results about the alternative models, showing the percentage of psd estimations in the cases with i.i.d., autocorrelated and heteroskedastic noise, are reported in appendix \ref{sec:appmod}. It is worth underlying, as final consideration, that the results in terms or RMISE always see the PDF estimator as the top performer, and also the ranking in terms of MISE remains unchanged in most of the cases, when moving to the alternative models.

\begin{table}[htbp!]
	\centering
		\begin{tabular}{c|c|c|c||c|c|c}
			\hline\hline
			Estimator & SV1F & SV2F & RH & SV1F & SV2F & RH \\
			\hline\hline
			& \multicolumn{3}{c||}{d=2} & \multicolumn{3}{c}{d=20}\\
			\hline
			PDF &  100\% &  100\% & 100\% & 100\% & 100\% & 100\% \\
			LMM &  100\% &  100\% & 100\% & 100\% &  100\% & 100\% \\
			STS &  100\% &  100\% & 100\% & 99.86\% & 99.86\% & 100\% \\
			\hline
			& \multicolumn{3}{c||}{d=5} & \multicolumn{3}{c}{d=25}\\
			\hline
			PDF & 100\% &  100\%  & 100\% & 100\% & 100\% &  100\% \\
			LMM & 100\% &  100\%  & 100\% & 100\% & 100\% &  100\% \\
			STS & 100\% &  100\%  & 100\% & 98.45\% & 97.20\% & 96.54\% \\
			\hline
			& \multicolumn{3}{c||}{d=10} & \multicolumn{3}{c}{d=30}\\
			\hline
			PDF & 100\% & 100\% & 100\% & 100\% & 100\% & 100\% \\
			LMM & 100\% & 100\% & 100\% & 100\% & 100\% & 100\% \\
			STS & 100\% & 100\% & 100\% & 94.25\% & 92.80& 29.72\% \\
			\hline
			& \multicolumn{3}{c||}{d=15} & \multicolumn{3}{c}{d=40}\\
			\hline
			PDF & 100\% & 100\% & 100\% & 100\% & 100\% & 100\% \\
			LMM & 100\% & 100\% & 100\% & 100\% & 100\% & 100\% \\
			STS & 100\% & 100\% & 100\% & 34.04\% & 32.12\% & 27.71\% \\
			\hline
		\end{tabular}
		\caption{\% of psd matrix produced by each estimator, when the efficient price process is produced by alternative models.}\label{tab:models}
\end{table}

\section{Conclusions}\label{sec:conc}

In the present wok a modified version of the classical Fourier estimator for spot covariance by \citet{malliavin2009fourier} has been proposed to overcame the difficulty of obtaining symmetric and positive semi-definite estimation of the spot variance-covariance matrix. The proposed estimator is know to be consistent (\citet{aka2023pdf}), and here we proved its positive semi-definiteness. A numerical study has been carried out to evaluate the optimal choice of the parameters $N$ and $M$ in a variety of settings. The optimal couple seems to be quite stable, and, as usual for the family of Fourier estimators, if the data is heavily contaminated by noise $N$ should be reduced. Based on those findings, a thorough simulation study has been carried out to evaluate the accuracy of the estimator and its actual ability in producing psd estimations. Comparing the results with the ones of two alternative estimators present in the literature, that are not proven to be positive semi-definite, we found out that the proposed PDF estimator usually outperforms the competitors in terms of mean square error., showing that is possible to obtain desirable properties of the estimated spot covariance matrix without reducing the accuracy of the estimation. Moreover, the proposed estimator is found to be robust to a variety of specification of market microstructure noise, via an appropriate choice of the cutting frequency $N$. The robustness of our results are confirmed using alternative data generating processes.

\begin{appendices}
\section{Proof of Theorem\ref{theo:pdf}}\label{sec:appprof}	

Let $a_j$ for $j=1,2,3$ be arbitrary functions on $\mathbb{Z}$, from the definitions of $\mathcal{K}$ and $\mathcal{S}(k)$ we notice that:

\begin{flalign*}
& \sum_{k \in \mathcal{K}} \sum_{(s,s') \in \mathcal{S}(k)} a_1(k) a_2(s) a_3(s')\\
& = \sum_{k=0}^{2N} \sum_{v=0}^{2N-k} a_1(k) a_2(-N+k+v) a_3(N-v) + \sum_{k=-2N}^{-1} \sum_{v=0}^{2N+k} a_1(k) a_2(N+k-v) a_3(-N+v) \eqqcolon A+B.
\end{flalign*}

For the first term we have:

\begin{align*}
 A	& = \sum_{k=0}^{2N} \sum_{u=k-N}^{N} a_1(k) a_2(k-u)  a_3(u) \\
 	& = \sum_{u=-N}^{N} \sum_{k=0}^{u+N} a_1(k) a_2(k-u)  a_3(u) \\
 	& = \sum_{u=-N}^{N} \sum_{u'=-N}^{N} a_1(u+u') a_2(u') a_3(u),
\end{align*}

where we set $u = N - v$ in the first line, changed the order of the
summations in the second line, and put $u' = k - u$. Similarly, using the convention that $\sum_{u=0}^{-1}=0$, for the second term we have:

\begin{align*}
	B	& = \sum_{k=-2N}^{-1} \sum_{u=-N}^{N+k} a_1(k) a_2(k-u)  a_3(u) \\
	& = \sum_{u=-N}^{N} \sum_{k=u-N}^{-1} a_1(k) a_2(k-u)  a_3(u) \\
	& = \sum_{u=-N}^{N} \sum_{u'=-N}^{-u-1} a_1(u+u') a_2(u') a_3(u).
\end{align*}

Thus we see that
\[
\sum_{k \in \mathcal{K}} \sum_{(s,s') \in \mathcal{S}(k)} a_1(k) a_2(s) a_3(s') = \sum_{u=-N}^{N}\sum_{u'=-N}^{N}a_1(u+u') a_2(u') a_3(u).
\]

When $a_1(k)=c(k)e^{ikt}$, $a_2(s)=e^{-ist^{j'}_{l'}}$ and $a_3(s')=e^{-ist^j_l}$, using the change of variable $ u- \rightarrow -u'$, we obtain:

\[
\hat{V}^{j,j'}_N(t)= \sum_{l=1}^{n_j}\sum_{l'=1}^{n_{j'}}\sum_{u=-N}^{N}\sum_{u'=-N}^{N}c(u-u')e^{ i u(t-t_l^j)}e^{- i u'(t-t_{l'}^{j'})} \Delta(X_l^j)\Delta(X_{l'}^{j'}).
\]

Then, for $x \in \mathbf{C}^d$

\begin{align*}
& \sum_{j,j'} \hat{V}^{j,j'}_N(t)x_j \overline{x_{j'}}\\
& = \sum_{u=-N}^{N}\sum_{u=-N}^{N}c(u-u')\left( \sum_{j=1}^dx_j\sum_{l=1}^{n_j}e^{iu(t-t^j_l)}\Delta X^j_l\right) \left(\sum_{j'=1}^dx_j\sum_{l'=1}^{n_{j'}}e^{-iu(t-t^{j'}_{l'})}\Delta X^{j'}_{l'}\right) \\
& =\sum_{u=-N}^{N}\sum_{u'=-N}^{N}c(u-u') f(u) \overline{f(u')} \ge0.
\end{align*}

with $f(u) \coloneqq \sum_{j=1}^dx_j \sum_{l=1}^{n_j}e^{iu(t-t^j_l)}\Delta X^j_l$. The proof is complete.

\section{Additional results of comparison for alternative models}\label{sec:appmod}

\begin{table}[H]
	\hspace{-2cm}
	\small
	\begin{tabular}{c|c|c|c||c|c|c||c|c|c||c|c|c}
		\hline\hline
		Estimator & SV1F & SV2F & RH & SV1F & SV2F & RH & SV1F & SV2F & RH & SV1F & SV2F & RH \\
		\hline\hline
		& \multicolumn{3}{c||}{d=5, $\sigma_{\eta}=1$} & \multicolumn{3}{c||}{d=5, $\sigma_{\eta}=1.5$} & \multicolumn{3}{c||}{d=5, $\sigma_{\eta}=2$} & \multicolumn{3}{c}{d=5, $\sigma_{\eta}=2.5$}\\
		\hline
		PDF &  100\%  &  100\%  & 100\%  & 100\%  & 100\% & 100\% & 100\% & 100\% & 100\% & 100\%& 100\% & 100\%\\
		LMM &  100\%  &  100\%  & 100\%  & 100\%  & 100\% &   100\% & 100\% & 99.85\% & 99.68\% & 97.28\%& 93.98\%& 94.98\%\\
		STS &  100\%  &  100\%  & 100\%  & 100\%  & 100\% &   100\% & 99.77\% & 99.77\% & 99.79\% & 97.95\%& 97.33\%& 97.70\%\\
		\hline
		& \multicolumn{3}{c||}{d=10, $\sigma_{\eta}=1$} & \multicolumn{3}{c||}{d=10, $\sigma_{\eta}=1.5$} & \multicolumn{3}{c||}{d=10, $\sigma_{\eta}=2$} & \multicolumn{3}{c}{d=10, $\sigma_{\eta}=2.5$}\\
		\hline
		\hline
		PDF & 100\% &  100\%  &  100\% &  100\% &  100\% & 100\% & 100\% & 100\% & 100\% & 100\%& 100\%& 100\%\\
		LMM & 100\% &  100\%  & 100\% &  99.96\% &  100\% &  100\% & 99.70\%& 99.25\% &95.26& 90.03\%& 85.48\%& 87.11\%\\
		STS & 100\% &  100\%  & 100\% & 99.60\% & 99.69\% &  99.54\%  & 90.91\% & 89.16\% &91.10& 52.28\%& 49.62\%& 54.59\%\\
		\hline
		& \multicolumn{3}{c||}{d=15, $\sigma_{\eta}=1$} & \multicolumn{3}{c||}{d=15, $\sigma_{\eta}=1.5$} & \multicolumn{3}{c||}{d=15, $\sigma_{\eta}=2$} & \multicolumn{3}{c}{d=15, $\sigma_{\eta}=2.5$}\\
		\hline
		\hline
		PDF & 100\% & 100\% & 100\% & 100\% & 100\% & 100\% & 100\% & 100\%& 100\% & 100\% & 100\%& 100\%\\
		LMM & 100\% & 100\% & 100\% & 99.96\% & 99.80\% & 99.80\% & 99.25\%& 97.85\%&  98.02\% & 79.28\%& 75.16\%& 77.98\%\\
		STS & 100\% & 99.90\% & 99.96\% & 90.52\% & 88.66\%& 90.81\%& 37.21\% & 29.88\%&  39.24\% &23.72\%& 15.64\% & 18.25\%\\
		\hline
		& \multicolumn{3}{c||}{d=20, $\sigma_{\eta}=1$} & \multicolumn{3}{c||}{d=20, $\sigma_{\eta}=1.5$} & \multicolumn{3}{c||}{d=20, $\sigma_{\eta}=2$} & \multicolumn{3}{c}{d=20, $\sigma_{\eta}=2.5$}\\
		\hline
		\hline
		PDF & 100\% & 100\% & 100\% & 100\% & 100\% & 100\% & 100\%& 100\% & 100\% & 100\%& 100\%& 100\%\\
		LMM & 100\% & 100\% & 100\% & 99.96\% & 98.27\% & 98.82\% & 97.95\%& 95.23\% & 95.88\% & 65.91\%& 63.84\%& 65.95\%\\
		STS & 97.37\% & 96.68\% & 97.14\% & 42.42\% & 49.39\%& 44.44\%& 28.56\% & 22.64\% & 25.84\% & 1.80\%& 0.28\%& 0.02\%\\
		\hline
	\end{tabular}
	\caption{\% of psd matrix produced by each estimator, when the efficient price process is produced by alternative models, in presence of i.i.d. noise.}
\end{table}

\begin{table}[H]
	\small
	\begin{tabular}{c|c|c|c||c|c|c||c|c|c}
		\hline\hline
		Estimator & SV1F & SV2F & RH & SV1F & SV2F & RH & SV1F & SV2F & RH\\
		\hline\hline
		& \multicolumn{3}{c||}{d=5, $\theta_{\eta}=0.2$ } & \multicolumn{3}{c||}{d=5, $\theta_{\eta}=0.3$} & \multicolumn{3}{c}{d=5, $\theta_{\eta}=0.4$}\\
		\hline
		PDF &  100\%  &  100\%  & 100\%  & 100\%  & 100\% & 100\% & 100\% & 100\% & 100\% \\
		LMM &  100\%  &  99.96\%  & 100\%  & 100\%  & 100\% &   100\% & 100\% & 100\% & 100\% \\
		STS &  97.39\%  &  97.20\%  & 97.83\%  & 99.21\%  & 99.14\% &   99.83\% & 99.77\% & 99.25\% & 99.73\%\\
		\hline
		& \multicolumn{3}{c||}{d=10, $\theta_{\eta}=0.2$ } & \multicolumn{3}{c||}{d=10, $\theta_{\eta}=0.3$} & \multicolumn{3}{c}{d=10, $\theta_{\eta}=0.4$}\\
		\hline
		\hline
		PDF & 100\% &  100\%  &  100\% &  100\% &  100\% & 100\% & 100\% & 100\% & 100\% \\
		LMM & 99.85\% &  98.20\%  & 100\% &  98.16\% &  96.90\% &  100\% & 100\%& 100\% &100\%\\
		STS & 49.23\% &  47.87\%  & 53.55\% & 73.25\% & 71.10\% &  73.25\%  & 90.39\% & 88.34\% & 89.65\%\\
		\hline
		& \multicolumn{3}{c||}{d=15, $\theta_{\eta}=0.2$ } & \multicolumn{3}{c||}{d=15, $\theta_{\eta}=0.3$} & \multicolumn{3}{c}{d=15, $\theta_{\eta}=0.4$}\\
		\hline
		\hline
		PDF & 100\% & 100\% & 100\% & 100\% & 100\% & 100\% & 100\% & 100\%& 100\% \\
		LMM & 99.12\% & 97.90\% & 96.55\% & 99.85\% & 99.73\% & 98.16\% & 99.88\%& 95.36\%&  97.22\% \\
		STS & 10.22\% & 10.63\% & 9.36\% & 11.32\% & 15.38\%& 14.64\%& 21.32\% & 20.44\%&  20.89\% \\
		\hline
		& \multicolumn{3}{c||}{d=20, $\theta_{\eta}=0.2$ } & \multicolumn{3}{c||}{d=20, $\theta_{\eta}=0.3$} & \multicolumn{3}{c}{d=20, $\theta_{\eta}=0.4$}\\
		\hline
		\hline
		PDF & 100\% & 100\% & 100\% & 100\% & 100\% & 100\% & 100\%& 100\% & 100\% \\
		LMM & 98.52\% & 97.50\% & 96.55\% & 98.63\% & 99.20\% & 98.25\% & 99.85\%& 96.17\% & 96.82\% \\
		STS & 1.83\% & 1.32\% & 3.74\% & 4.57\% & 3.22\%& 4.72\% & 8.94\% & 8.55\% & 9.33\% \\
		\hline
	\end{tabular}
	\caption{\% of psd matrix produced by each estimator, when the efficient price process is produced by alternative models, in presence of autocorrelated noise.}
\end{table}

\begin{table}[H]
	\small
	\begin{tabular}{c|c|c|c||c|c|c||c|c|c}
		\hline\hline
		Estimator & SV1F & SV2F & RH & SV1F & SV2F & RH & SV1F & SV2F & RH\\
		\hline\hline
		& \multicolumn{3}{c||}{d=5,  $\bar{\sigma}_\eta=3$} & \multicolumn{3}{c||}{d=5, $\bar{\sigma}_\eta=3.5$} & \multicolumn{3}{c}{d=5, $\bar{\sigma}_\eta=4$}\\
		\hline
		PDF &  100\%  &  100\%  & 100\%  & 100\%  & 100\% & 100\% & 100\% & 100\% & 100\% \\
		LMM &  100\%  &  99.96\%  & 99.96\%  & 99.25\%  & 98.95\% &   99.18\% & 92.79\% & 91.65\% & 91.31\% \\
		STS &  89.60\%  &  82.73\%  & 84.52\%  & 72.26\%  & 72.97\% &   74.06\% & 61.17\% & 60.53\% & 63.01\%\\
		\hline
		& \multicolumn{3}{c||}{d=10,  $\bar{\sigma}_\eta=3$} & \multicolumn{3}{c||}{d=10, $\bar{\sigma}_\eta=3.5$} & \multicolumn{3}{c}{d=10, $\bar{\sigma}_\eta=4$}\\
		\hline
		\hline
		PDF & 100\% &  100\%  &  100\% &  100\% &  100\% & 100\% & 100\% & 100\% & 100\% \\
		LMM & 99.85\% &  99.73\%  & 99.77\% &  95.78\% &  95.74\% &  95.61\% & 79.97\%& 80.03\% &81.27\%\\
		STS & 45.15\% &  44.82\%  & 45.95\% & 36.34\% & 36.11\% &  36.38\%  & 30.33\% & 29.74\% &30.45\%\\
		\hline
		& \multicolumn{3}{c||}{d=15,  $\bar{\sigma}_\eta=3$} & \multicolumn{3}{c||}{d=15, $\bar{\sigma}_\eta=3.5$} & \multicolumn{3}{c}{d=15, $\bar{\sigma}_\eta=4$}\\
		\hline
		\hline
		PDF & 100\% & 100\% & 100\% & 100\% & 100\% & 100\% & 100\% & 100\%& 100\% \\
		LMM & 99.12\% & 98.02\% & 99.25\% & 90.68\% & 89.58\% & 91.22\% & 67.04\%& 66.54\%&  68.96\% \\
		STS & 30.58\% & 29.88\% & 30.81\% & 23.92\% & 23.39\%& 23.66\%& 19.40\% & 19.32\%&  18.94\% \\
		\hline
		& \multicolumn{3}{c||}{d=20,  $\bar{\sigma}_\eta=3$} & \multicolumn{3}{c||}{d=20, $\bar{\sigma}_\eta=3.5$} & \multicolumn{3}{c}{d=20, $\bar{\sigma}_\eta=4$}\\
		\hline
		\hline
		PDF & 100\% & 100\% & 100\% & 100\% & 100\% & 100\% & 100\%& 100\% & 100\% \\
		LMM & 97.58\% & 97.09\% & 97.58\% & 83.06\% & 80.64\% & 83.44\% & 52.92\%& 51.88\% & 55.24\% \\
		STS & 21.37\% & 21.37\% & 21.30\% & 15.77\% & 15.66\%& 15.31\%& 11.09\% & 11.05\% & 10.99\% \\
		\hline
	\end{tabular}
	\caption{\% of psd matrix produced by each estimator, when the efficient price process is produced by alternative models, in presence of heteroskedastic noise.}
\end{table}

\end{appendices}

\printbibliography

\end{document}